\documentclass[3p]{elsarticle}
\usepackage{amsmath}
\usepackage{amsthm}
\usepackage{amssymb}
\usepackage[usenames,dvipsnames]{color}
\usepackage{enumitem}
\usepackage{tikz}
\tikzset{cross/.style={cross out, draw=black, minimum size=2*(#1-\pgflinewidth), inner sep=0pt, outer sep=0pt},
cross/.default={8pt}}

\usepackage{subfigure}
\usepackage[ruled,vlined,linesnumbered]{algorithm2e}
\usepackage{lineno}

\newcommand{\comm}[1] {}
\newcommand{\VD}{\mathcal{VD}}
\newcommand{\PP}{\mathcal{P}}
\newcommand{\PPP}{\texttt{P}}
\newcommand{\HH}{\mathcal{I}}
\newcommand{\CC}{\mathcal{C}}
\newcommand{\TT}{\mathcal{T}}
\newcommand{\PT}{\PP\TT}
\newcommand{\HT}{\HH\TT}
\newcommand{\FF}{\mathcal F}
\newcommand{\stepover}{\delta}
\newcommand{\tStepover}{\Delta}
\newcommand{\parentEdge}{\texttt{ParentE}}
\newcommand{\childEdges}{\texttt{ChildEs}}
\newcommand{\nodeStart}{\texttt{Start}}
\newcommand{\nodeEnd}{\texttt{End}}
\newcommand{\indexInParentNode}{\texttt{IndexInStart}}
\newcommand{\getNextEdge}{\texttt{NextCCW}}

\newcommand{\ttimeO}{\texttt{TmNd}}
\newcommand{\ttimeT}{\texttt{TmEd}}
\newcommand{\speedO}{\texttt{VeNd}}
\newcommand{\speedT}{\texttt{VeEd}}
\newcommand{\height}{\texttt{Hgt}}
\newcommand{\holeHeight}{\texttt{Isl}\height}
\newcommand{\boundaryHeight}{\texttt{Bnd}\height}
\newcommand{\getSpeed}{\texttt{GetSp}}
\newcommand{\getDist}{\texttt{GetDs}}
\newcommand{\getTime}{\texttt{GetTm}}
\newcommand{\getPoint}{\texttt{GetPt}}
\newcommand{\setTimesAndSpeeds}{\texttt{SetTmAndSp}}
\newcommand{\setAllTimesAndSpeeds}{\texttt{SetAllTmAndSp}}
\newcommand{\reuseAcc}{\texttt{ReuseAcc}}
\newcommand{\wavefront}{\texttt{Wf}}
\newcommand{\holeWavefront}{\texttt{Isl}\wavefront}
\newcommand{\boundaryWavefront}{\texttt{Bnd}\wavefront}
\newcommand{\wavefrontLength}{\wavefront\texttt{Lng}}
\newcommand{\totalWavefrontLength}{\texttt{Ttl}\wavefrontLength}
\newcommand{\rootNode}{\texttt{Root}}
\newcommand{\point}{\texttt{Pt}}
\newcommand{\spiral}{\texttt{Sp}}
\newcommand{\parent}{\texttt{Pa}}
\newcommand{\parentWavefrontCorner}{\parent\wavefront}
\newcommand{\parentSpiralCorner}{\parent\spiral}
\newcommand{\arc}{\texttt{Arc}}
\newcommand{\inflDist}{\texttt{InfDst}}
\newcommand{\weight}{\texttt{Wgt}}
\newcommand{\bridges}{\texttt{Bridges}}
\newcommand{\makeBridges}{\texttt{Make}\bridges}
\newcommand{\ora}[1]{\overrightarrow{#1}}
\newcommand{\picc}[1]{}

\journal{arXiv.org}

\begin{document}

\begin{frontmatter}

\title{Spiral Toolpaths for High-Speed Machining of 2D Pockets with or without Islands}
\author{Mikkel Abrahamsen\fnref{fn1}}
\fntext[fn1]{A large part of this work was made while the author worked for
Autodesk, Inc.}
\address{Department of Computer Science\\
University of Copenhagen\\
Universitetsparken 5\\
DK-2100 K\o{}benhavn \O\\
Denmark\\
\texttt{miab@di.ku.dk}}

\begin{abstract}
We describe new methods for the construction of spiral toolpaths for high-speed machining.
In the simplest case, our method takes a polygon as input and a number $\stepover>0$ and
returns a spiral starting at a central point in the polygon, going around towards the boundary
while morphing to the shape of the polygon. The spiral consists of
linear segments and circular arcs, it is $G^1$ continuous, it has no self-intersections, and
the distance from each point on the spiral to each of the neighboring revolutions is at most $\stepover$.
Our method has the advantage over previously described methods that it is
easily adjustable to the case where there is an island in the polygon to be avoided by the spiral.
In that case, the spiral starts at the island and morphs
the island to the outer boundary of the polygon. It is shown how to apply that method to make
significantly shorter spirals in polygons with no islands than what is obtained by conventional
spiral toolpaths. Finally, we show how to make
a spiral in a polygon with multiple islands by connecting the islands into one island.
\end{abstract}


\end{frontmatter}



\section{Introduction}

A fundamental problem often arising in the CAM\comm{\footnote{Computer-aided manufacturing.}} industry
is to find a suitable toolpath for milling a pocket that is defined by a shape in the plane.
A CNC\comm{\footnote{Computer numerical control.}} milling machine is programmed to follow
the toolpath and thus cutting a cavity with the shape of the given pocket in a solid piece of material.
The cutter of the machine can be regarded as a circular disc with radius $r$, and
the task is to find a toolpath in the plane
such that the swept volume of the disc, when the disc center is moved along the path,
covers the entire pocket. We assume for simplicity that the toolpath is allowed
to be anywhere in the pocket.

Some work has been made on spiral toolpaths that morphs a point
within the pocket to the boundary of the pocket \cite{bieterman2003, chuang2007, held2009, banerjee2012, held2014, huertas2014}.
The method described by Held and Spielberger
\cite{held2009} yields a toolpath that (i) starts at a user-specified
point within the pocket, (ii) ends when the boundary is reached, (iii) makes the cutter remove
all material in the pocket, (iv) has no self-intersections, (v)
is $G^1$ continuous\footnote{A plane curve
is \emph{$G^1$ continuous} or \emph{tangent continuous}
if there exists a continuous and differentiable parameterization of the curve.},
(vi) makes the width of material cut away at most $\stepover$ at any time, where
$\stepover$ is a user-defined constant called the \emph{stepover}.
We must have $\stepover<r$, since otherwise some material might not be cut away.
See figure \ref{spiral} for an example of such a spiral toolpath. It is the result of an algorithm
described in the present paper, but has similar appearance as the spirals described by Held and Spielberger
\cite{held2009}.

Most traditional toolpath patterns have many places where the cutter does not cut away any new
material, for instance in retracts where it is lifted and moved in the air to another place
for further machining, or self-intersections of the toolpath, where the tool
does not cut away anything new when it visits a place for the second time.
That may increase machining time and lead to visible marks on the
final product. Spiral toolpaths have the advantage that the cutter is
cutting during all of the machining and that, at the same time,
the user can control the stepover.
Spiral toolpaths are particularly useful when doing high-speed machining, where the
rotational speed of the cutter and the speed with which it is moved along the
toolpath is higher than in conventional milling.
We refer to Held and Spielberger \cite{held2009} for a more detailed discussion of
the benefits of spiral toolpaths compared to various other toolpath patterns
and more information on CNC milling in general.

Bieterman and Sandstrom \cite{bieterman2003} and Huertas-Tal{\'o}n et al.~\cite{huertas2014}
give methods
for computing spiral toolpaths by solving elliptic partial differential equation
boundary value problems defined on the pocket. However, the methods only work
for star-shaped pockets\footnote{A polygon is \emph{star-shaped} if there
exists a point $p$ in the polygon such that the segment $pq$ is contained in the polygon
for every other point $q$ in the polygon.} \cite{held2009}.

We describe an alternative construction of spirals that also satisfy the previously mentioned properties
of the construction of Held and Spielberger \cite{held2009}.
They define a wave that starts at a point in the Voronoi diagram of the pocket at time
$0$. When the time increases, the wave moves towards the boundary of the
pocket in every direction so that at time $1$,
it reaches the boundary everywhere. The shape of the wave at a certain time represents
the area machined at that time. Roughly speaking,
the spiral is obtained by traveling around the wave while the time
increases from $0$ to $1$.
Our method is similar up to this point, but the way we define the wave is different.
The wave defined in \cite{held2009} is the union of growing
disks placed on the Voronoi diagram of the pocket.
In our model, the wave is at any time a polygon with its corners on
the Voronoi diagram of the pocket.
Using that model, we define a spiral consisting of line segments which is
at last rounded by circular arcs to get a $G^1$ continuous curve.

In practice, it is very common that there is one or more islands in the
pocket that should be avoided by the cutter, for instance if there are areas of material that should
not be machined to the same depth.
It is only described by Held and Spielberger \cite{held2009} how to handle simply-connected
pockets, i.e., there must be no islands.
In their following paper \cite{held2014}, it is described how one can handle a pocket with an island
by connecting it to the boundary with a ``bridge'', effectively getting a pocket without islands.
Thus, the resulting spiral morphs a point to a shape consisting of the island,
the bridge, and the pocket boundary.
A big advantage of our method is that it has a natural extension to pockets with one
island in the sense that the spiral morphs the shape of the island to the pocket boundary.
We exploit the fact that the Voronoi diagram of a pocket with an island consists of exactly one cycle
and trees rooted at that cycle to extend our wave model to work in this case.
This is our biggest new contribution. We shall demonstrate natural applications of this method
to make significantly shorter spirals for pockets with no holes than one could obtain with previously described methods. We also show how to handle pockets with multiple holes.

The paper is based on the author's experiences while developing the morphed spiral strategy for the
Autodesk\texttrademark{} CAM products (HSMWorks,
Inventor HSM\textsuperscript{\textregistered}, and Fusion 360\textsuperscript{\textregistered}).
The morphed spiral seems to be quite popular among the users.
Due to the abundant number of real-world parts that have been available during the development,
we guarantee that it is possible
to make an efficient industrial-strength implementation of the algorithms described here.

We use Held's \texttt{VRONI} library for the computation of Voronoi diagrams \cite{held2001}.
All figures in the paper are automatically generated using our implementation of the algorithms.

The rest of the paper is structured as follows:
In section \ref{withoutHoles}, we describe our basic method for making a spiral
that morphs a point to the boundary in a simply-connected pocket.
Section \ref{pocketWithHole} describes how the method is adapted to a pocket with one island.
Using that method, we describe in section \ref{skeletonMethod} an alternative spiral in simply-connected pockets which will be superior to the one from section \ref{withoutHoles} in many cases.
In section \ref{multipleHoles}, we show how to construct at spiral
around arbitrarily many islands by first connecting the islands into one island.
Finally, we conclude the paper in section \ref{conclusion}
by suggesting some future paths of development of spiral
toolpaths.

\subsection{Notation and other general conventions}
We use zero-based numbering of arrays.
For a point $p=(x,y)$, the point $\widehat p=(-y,x)$ is the counterclockwise rotation of $p$.
Given two distinct points $p$ and $q$, $pq$ is the segment between $p$ and $q$
and $\ora{pq}$ is the half-line that starts at $p$ and contains $q$.
Given a set of points $S$ in the plane, $\partial S$ denotes the boundary of $S$.
In algorithms, we
use semicolon to separate different statements written on the same line.

\section{Computing a spiral in a pocket without islands}\label{withoutHoles}

\begin{figure*}
\centering
\subfigure[]{
\includegraphics{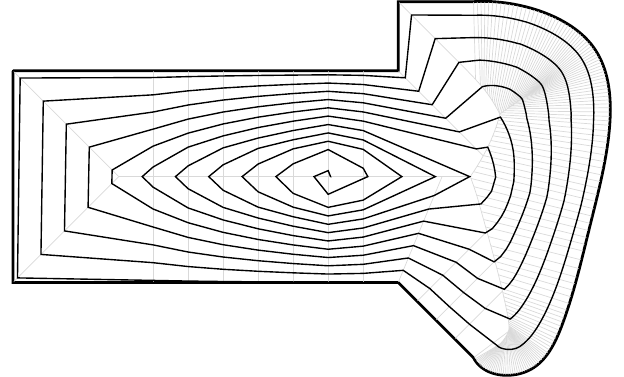}
\label{polylineSpiral}
}\quad
\subfigure[]{
\includegraphics{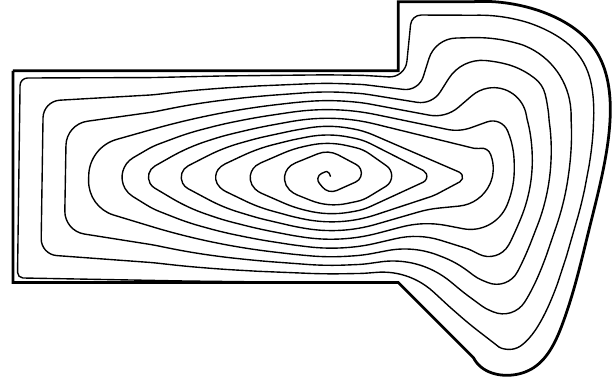}
\label{spiral}
}
\caption{
\subref{polylineSpiral} The polyline spiral and $\VD$ in gray.
\subref{spiral} The final rounded spiral.
}
\label{firstSpirals}
\end{figure*}

In this section we describe a method to compute a spiral in a given simply-connected
2D pocket $\PP$, see figure \ref{spiral}.
In practice, the boundary of a pocket is often described by line segments and more advanced pieces of
curves, such as like circular arcs, elliptic arcs, and splines.
However, it is always possible to use a sufficiently accurate linearization of the input,
so we assume for simplicity that $\PP$ is a polygon.

Our algorithm first constructs a polyline spiral, see figure \ref{polylineSpiral}.
The polyline spiral must respect the stepover $\delta$, i.e.,
the distance from every point to the neighboring revolutions
and the distance from the outermost revolution to $\partial\PP$
is at most $\delta$.
In section \ref{rounding}
we devise a method for rounding the polyline spiral to get a $G^1$ continuous spiral
consisting of line segments
and circular arcs. 

\begin{figure*}
\centering
\subfigure[] {
\includegraphics{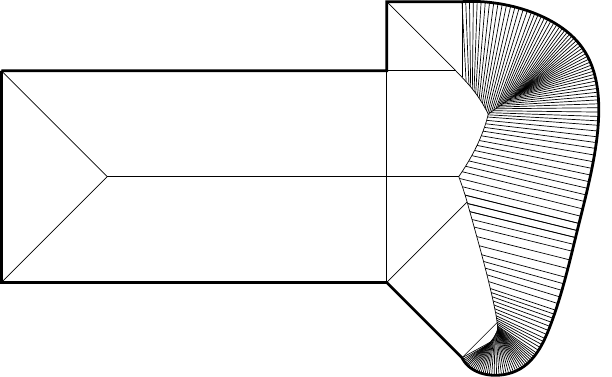}
\label{rawVD}
}\quad
\subfigure[]{
\includegraphics{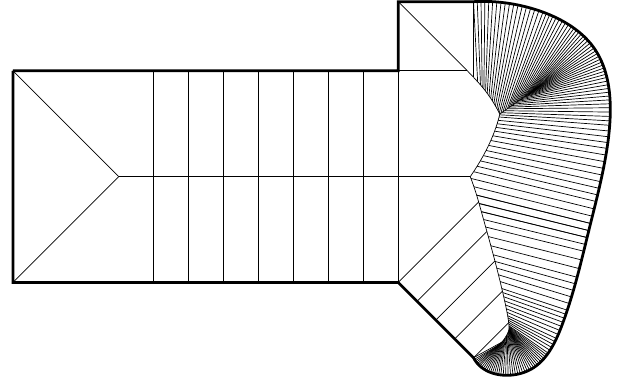}
\label{enrichedVD}
}
\subfigure[]{
\includegraphics{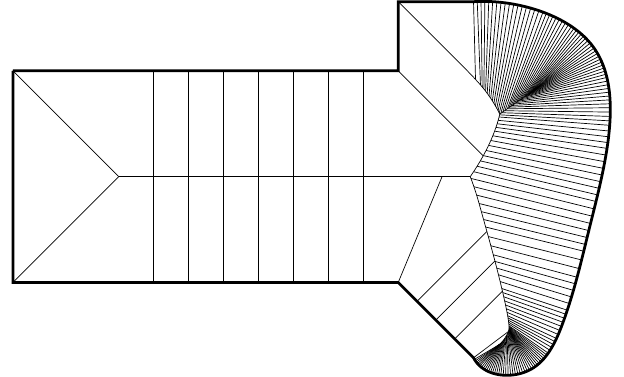}
\label{finalVD}
}
\caption{\subref{rawVD} The Voronoi diagram.
\subref{enrichedVD} The Voronoi diagram enriched with equidistantly placed
segments perpendicular to long edges.
\subref{finalVD} The final diagram $\VD$ where double edges going to concave corners
of $\PP$ are replaced by their angle bisector.
}
\end{figure*}

The corners of the
polyline spiral are points on the edges of the Voronoi diagram of $\PP$, and there
is a corner at each intersection point between the spiral and the Voronoi diagram.
We only consider the part of the Voronoi diagram inside $\PP$. We have found
that we get better results in practice by modifying the Voronoi diagram slightly.
We describe these modifications in sections \ref{enriching} and \ref{removing}
to avoid too many technical details here.
See figures \ref{rawVD}--\ref{finalVD} for a concrete example of the modifications
we make on the Voronoi diagram.
Let $\VD=\VD(\PP)$ be the modified Voronoi diagram of the pocket $\PP$. 
Like the Voronoi diagram of $\PP$, the modified diagram
$\VD$ has the following properties which are necessary and sufficient for
the computation of the spiral:
\begin{itemize}
\setlength\itemsep{0em}
\item $\VD$ is a plane tree contained in $\PP$,
\item each leaf of $\VD$ is on the boundary $\partial\PP$ of $\PP$,
\item there is at least one leaf of $\VD$ on each corner of $\PP$,
\item all the faces into which $\VD$ divides $\PP$ are convex.
\end{itemize}

\begin{figure*}
\centering
\subfigure[]{
\includegraphics{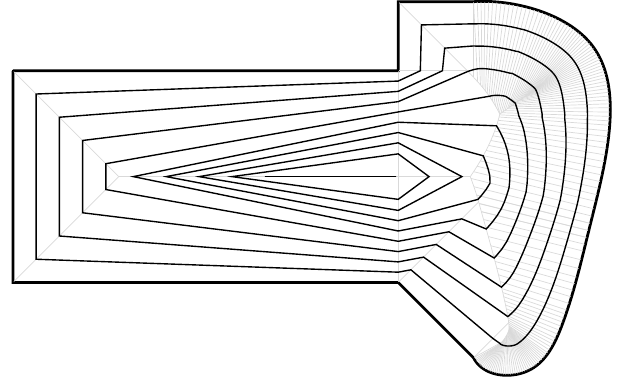}
\label{wavefrontsBad}
}\quad
\subfigure[]{
\includegraphics{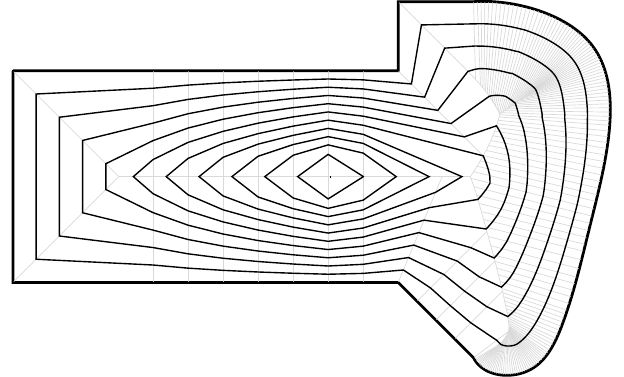}
\label{wavefrontsGood}
}
\caption{The wavefronts in the same polygon $\PP$ for the same stepover $\stepover$
but using two different diagrams to define the wavefronts.
\subref{wavefrontsBad} The wavefronts obtained using the Voronoi diagram. The Voronoi
diagram is in gray.
\subref{wavefrontsGood} The wavefronts obtained using $\VD$. $\VD$ is in gray.
}
\label{wavefronts}
\end{figure*}

\subsection{The wave model}\label{waveModel}

We imagine that a wave starts at time $t=0$ at the point $p_0$ inside $\PP$. The wave moves out
in every direction such that at time $t=1$, it has exactly the same shape as $\partial\PP$.
The shape of the wave at a specific time is called a \emph{wavefront}.
The wave is growing in the sense that if $0\leq t_1\leq t_2\leq 1$,
the wavefront at time $t_1$ is contained in the wavefront at time $t_2$.
We choose $p_0$ as a point in the diagram $\VD$ and consider $\VD$ as a tree
rooted at $p_0$. We define the time
at which the wave hits each node and the speed with which it travels on each edge in $\VD$.
The speed of the wave is always constant or decreasing.
Thus, we create a continuous map $\theta:\VD\mapsto [0,1]$ that assigns
a time value between 0 and 1 to each point on $\VD$. If $p$ is a point moving along a path
on $\VD$ from $p_0$ to any leaf,
the value $\theta(p)$ increases monotonically from 0 to 1.
For each time $t\in[0,1]$, the wavefront is a polygon inside $\PP$ and the vertices of the wavefront
are all the points $p$ on $\VD$ such that $\theta(p)=t$. Note that there is exactly one such point
on each path from $p_0$ to a leaf of $\VD$ for a given $t\in [0,1]$.

We define a time step $\tStepover = 1/r$ for some
integer $r$ and compute the wavefront at the times
$t\in \{0,\tStepover, 2\tStepover, \ldots, r\tStepover\}$, where $r\tStepover=1$,
see figure \ref{wavefronts}.
By \emph{wavefront} $i$, we mean the wavefront at time $i\tStepover$.
We choose $r$ such that the distance from each point on
wavefront $i$ to each of the wavefronts $i-1$ and $i+1$
is at most $\stepover$ when $i > 0$ and $i<r$, respectively.
In other words, the \emph {Hausdorff distance} between
two neighboring wavefronts is at most $\stepover$.
Recall that the Hausdorff distance between two sets $A$ and $B$ is
$\max\{d(A,B),d(B,A)\}$, where $d(A,B)=\max_{a\in A} \min_{b\in B} \lVert a-b\rVert$.
For each $i=1,\ldots, r$,
we compute a revolution of the polyline spiral by interpolating between the wavefronts $i-1$
and $i$. We describe in sections \ref{constructingWavefronts}
and \ref{interpolating} how to make the wavefronts and the interpolation such
that the stepover is respected between neighboring revolutions.

\subsection{Choosing the starting point $p_0$ and the number of revolutions of the spiral}\label{startingPoint}

In order to get a spiral with small length, we try to minimize the number of revolutions.
Consider the longest path from $p_0$ to a leaf in $\VD$.
The length of a path is the sum of edge lengths on the path.
If $h$ is the length of the longest path, then $\lceil h/\stepover\rceil+1$ wavefronts
are necessary and sufficient for the stepover to be respected between all neighboring wavefronts.
Therefore, we choose $p_0$ as the point in $\VD$ that minimizes the longest distance to a leaf in
$\VD$. That is a unique point traditionally known as the \emph{center} of $\VD$.
Handler \cite{handler1973} gives a simple algorithm to compute $p_0$ linear time in the size of
$\VD$.
The center will most likely not be a node in $\VD$, but an interior point on some edge. In that
case, we split the edge into two edges by introducing a node at $p_0$.

\subsection{Our representation of $\VD$}

We consider $\VD$ as a directed, rooted tree with the node $\rootNode$ at $p_0$ being the root.
We let $\point[n]$ be the position of the node $n$. Let
$\VD[n]$ be the subtree rooted at node $n$.
We store a pointer $\parentEdge[n]$ to the edge having end node $n\neq \rootNode$.
We say that edge $\parentEdge[n]$ is the \emph{parent edge} of node $n$ and any edge
having start node $n$.
We also store an array
$\childEdges[n]$ of the edges going out of $n$ sorted in counterclockwise order with the
edge following $\parentEdge[n]$ being the first. For $\rootNode$, the choice of the first child
edge does not matter.
For each edge $e$, we store pointers $\nodeStart[e]$ and $\nodeEnd[e]$ to 
the start and end nodes of $e$. We also store an index $i = \indexInParentNode[e]$
such that $\childEdges[\nodeStart[e]][i]=e$. If $e$ is an edge, we say that $\nodeStart[e]$
and $\nodeEnd[e]$ are \emph{incident} to $e$ and that $e$ is \emph{incident} to
$\nodeStart[e]$ and $\nodeEnd[e]$.
For an edge $e_1$ and node $n$ incident to $e_1$,
we let $\getNextEdge(e_1,n)=e_2$,
where $e_2$ is the edge after $e_1$ among the edges incident to $n$ in counterclockwise
order. The function $\getNextEdge$ can be implemented so that it runs in constant time
using the values defined here.

Using $\getNextEdge$, we can traverse all of $\VD$ in counterclockwise direction in linear time.
We start setting $(n,e)=(\rootNode,\childEdges[\rootNode][0])$.
In each iteration, we let $n$ be the other node incident to $e$ and
then set $e=\getNextEdge(e,n)$.
We stop when we have traversed every edge, i.e., when
$(e,n)=(\rootNode,\childEdges[\rootNode][0])$ at the end of an iteration. Note that each edge $e$
is visited twice, once going down the tree $\VD[\nodeStart[e]]$ and once going up.

\subsection{Defining the movement of the wave}\label{movementOfWave}

Let $\height[n]$ for each node $n$ be the maximal distance from $n$ to a leaf in $\VD[n]$.
All the $\height$ values can be computed in linear time by traversing $\VD$ once.
For each node $n$, we define the time $\ttimeO[n]$ where the wave reaches $n$.
We set $\ttimeO[\rootNode]=0$. We also define the speed $\speedO[n]$ that the wave has when it
reaches $n$. We set $\speedO[\rootNode]=\height[\rootNode]$.
The wave starts at the root at time $t=0$
and travels with constant speed $\speedO[\rootNode]$ on the paths to the farthest leafs in $\VD$.
(Due to our choice of the starting point $p_0$, there will always be at least two paths
from $p_0$ to a leaf with maximum length.)
Hence,
it reaches those leafs at time $t=1$. On all the shorter paths, we make the wave slow down
so that it reaches every leaf at time $t=1$.
In \ref{appA}, we describe in detail a possible way of defining the speed and time of the wave
on each point in $\VD$.

When the movement of the wave is defined,
we may define $\getPoint(e,t)$ as the point on edge $e$ with $\theta(\getPoint(e,t))=t$,
where $\ttimeO[\nodeStart[e]]\leq t\leq \ttimeO[\nodeEnd[e]]$.

\subsection{Constructing the wavefronts}\label{constructingWavefronts}

\begin{figure*}[h]
\centering
\subfigure[]{
\includegraphics[scale=0.97]{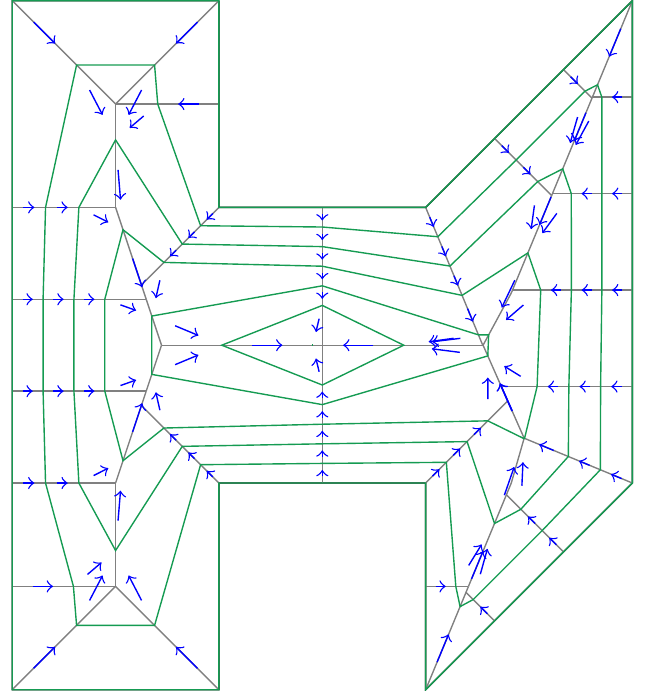}
\label{wavefrontsParents}
}\quad
\subfigure[]{
\includegraphics[scale=0.97]{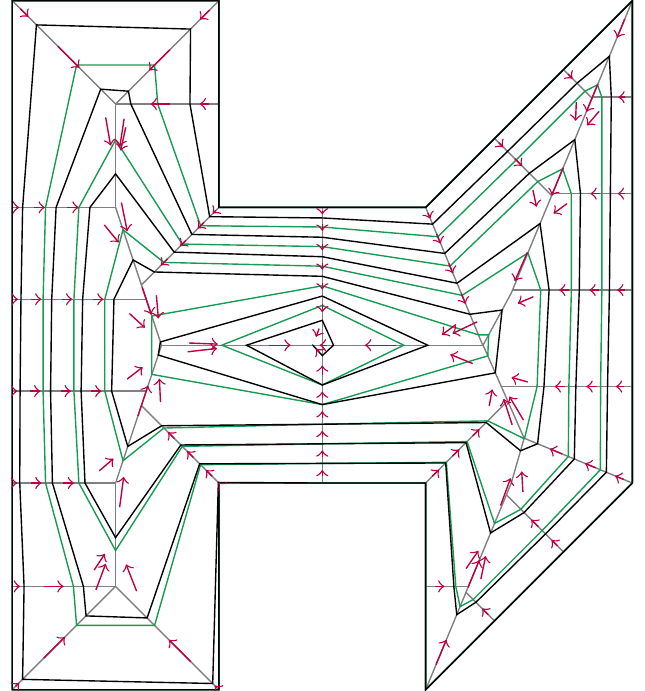}
\label{polylineSpiralParents}
}
\caption{The construction of a polyline spiral in a polygon $\PP$:
\subref{wavefrontsParents} The wavefronts in green and blue arrows from each wavefront corner
$\wavefront[i][w]$
to its parent $\wavefront[i-1][\parentWavefrontCorner[i][w]]$.
The diagram $\VD$ is in gray.
\subref{polylineSpiralParents} The polyline spiral in black obtained by interpolating
between the wavefronts. The purple arrows are from each corner $\spiral[i]$
of the spiral to its parent $\spiral[\parent[i]]$.
}
\label{polylineSpiralFig}
\end{figure*}

We make a spiral with $r=\lceil \frac {\height[\rootNode]}{\stepover'}\rceil$ revolutions, where
$\stepover'=0.95\cdot \stepover$.
We use the slightly smaller stepover $\stepover'$
so that the maximum distance between two neighboring
revolutions is smaller. That gives more flexibility to smooth the spiral later on as described in section \ref{rounding}.
We set $\tStepover=1/r$ and compute a wavefront for each of the times
$\{0,\tStepover,2\tStepover,\ldots,r\tStepover\}$. The two-dimensional array
$\wavefront$ stores the wavefronts, so that the wavefront at time $i\tStepover$ is the array
$\wavefront[i]$. Wavefront $i$ is constructed by traversing $\VD$ once and finding every
point on $\VD$ with time $i\tStepover$ in counterclockwise order.
Let $e$ be an edge we have not visited before, and let $n=\nodeStart[e]$ and
$m=\nodeEnd[e]$. There is a corner of wavefront $i$ on $e$ if
$\ttimeO[n] < i\tStepover \leq\ttimeO[m]$. If that is the case, we add
$\getPoint(e,i\tStepover)$ to the end of $\wavefront[i]$.
We make one corner in $\wavefront[0]$ for each of the child edges of the root,
$\childEdges[\rootNode]$, and all these corners are copies of
the point $\point[\rootNode]$. Using this construction,
there is exactly one corner of each wavefront on each path from
$\rootNode$ to a leaf of $\VD$.

For each corner $\wavefront[i][w]$, we store the length of the part of the
wavefront up to the corner, i.e.~$\wavefrontLength[i][0]=0$ and
$$
\wavefrontLength[i][w]=\sum_{j=1}^{w} \lVert \wavefront[i][j]-\wavefront[i][j-1]\rVert,
$$
for $w\geq 1$. Here $\lVert\cdot\rVert$ is the Euclidean norm. We also store the total length of
$\wavefront[i]$ as $\totalWavefrontLength[i]$.

We introduce a rooted tree with the wavefront corners as the nodes.
See figure \ref{wavefrontsParents}.
The parent of a corner $\wavefront[i][w]$, $i>0$, is the
unique corner $\wavefront[i-1][pw]$ on wavefront $i-1$ on the path from
$\wavefront[i][w]$ to $\rootNode$.
We store $pw$ as $\parentWavefrontCorner[i][w]$, i.e.,
the parent of $\wavefront[i][w]$ is $\wavefront[i-1][\parentWavefrontCorner[i][w]]$.

Since the distance from each wavefront corner to each of its children is at most $\stepover'$,
the Hausdorff distance between two neighboring wavefronts is also at most $\stepover'$.
Furthermore, since the wave is moving with positive speed towards the leafs of $\VD$ and the faces
into which $\VD$ divides $\PP$ are convex, neighboring wavefronts do not intersect each other.
From the order in which the corners of a wavefront is constructed, it is also clear that
a wavefront does not intersect itself.

\subsection{Interpolating between the wavefronts}\label{interpolating}

We construct a polyline spiral stored as an array $\spiral$. For each $i=1,\ldots, r$, we construct one
revolution of the spiral by interpolating between wavefront $i-1$ and wavefront $i$. Every
corner of the spiral is a point on $\VD$. There is exactly one spiral corner
on the path in $\VD$ from each wavefront corner $\wavefront[i][w]$ to its parent
$\wavefront[i-1][\parentWavefrontCorner[i][w]]$.
Assume for now that we know how to choose the actual corners of the polyline spiral. We shall get
back to this shortly.

The first corner $\spiral[0]$ is on the root node of $\VD$, and for every other
corner $\spiral[s]$, $s>0$, we store an index of the parent $\parent[s]$,
such that the parent $\spiral[\parent[s]]$
is the first corner we meet on the path in $\VD$ from $\spiral[s]$ to the root.
Figure \ref{polylineSpiralParents} shows the resulting polyline spiral and the parents of each corner.
We define the spiral
such that the distance between a spiral corner and its parent is at most $\stepover'$. It follows that
the distance from a point on the polyline spiral to the neighboring revolutions is at most $\stepover'$.

Here we describe how to define the $\parent$-pointers.
When we have constructed
a spiral corner $\spiral[s]$ which is on the path from $\wavefront[i][w]$ to its parent wavefront
corner $\wavefront[i-1][pw]$, we know that the first spiral corner on the path from
$\wavefront[i][w]$ to the root is $\spiral[s]$, and we store this information as
$\parentSpiralCorner[i][w]=s$.
Therefore, when we have made a new spiral corner $\spiral[r]$ on the path from $\wavefront[i+1][w']$
to its parent $\wavefront[i][pw']$, the parent of $\spiral[r]$ is defined to be
$\parent[r]=\parentSpiralCorner[i][pw']$. By doing so, the corner
$\spiral[\parent[r]]$ will be
the first spiral corner on the path from $\spiral[r]$ to the root.

\begin{figure*}[h]
\centering
\subfigure[]{
\includegraphics{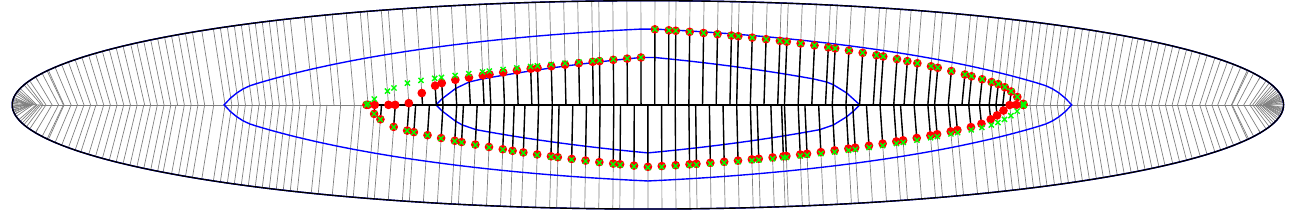}
\label{interpolation}
}\quad
\subfigure[]{
\includegraphics{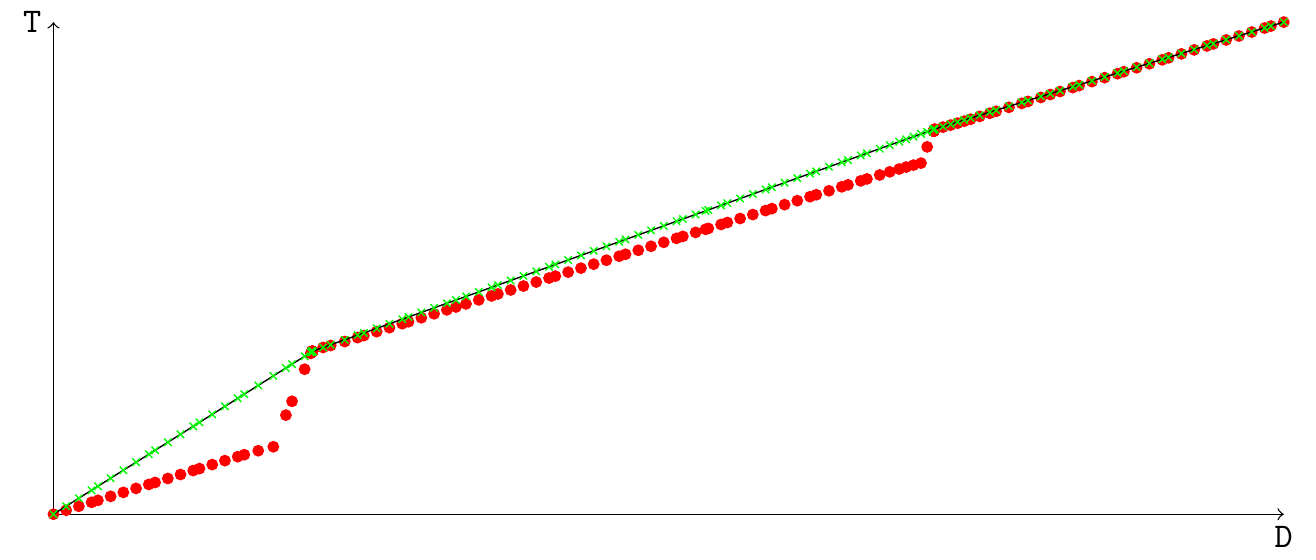}
\label{convexification}
}
\caption{\subref{interpolation} The interpolation between the two blue wavefronts. The boundary
$\PP$ is in black and the diagram $\VD$ is in gray. The red circles
are the points $\texttt Q[w]$ and the marked part of of $\VD$ is in black. The green crosses
are the resulting points of the polyline spiral stored in $\spiral$ after the convexification process.
\subref{convexification} Related values for the same interpolation:
The points $(\texttt D[w],\texttt T[w])$ are red circles.
The upper convex hull $\FF$ of the points is a black curve, and the green crosses are the points
$(\texttt D[w],\FF(\texttt D[w]))$ on that hull.}
\label{interpolationDetail}
\end{figure*}

In the following we describe how to choose the corners of the polyline spiral.
We assume that we have finished the revolution of the spiral between wavefronts $i-2$ and
$i-1$ and we show how to make the revolution between wavefronts $i-1$ and $i$.
For each wavefront corner $\wavefront[i][w]$, we find the point $\texttt{Q}[w]$ on the path to
$\wavefront[i-1][pw]$, where $pw=\parentWavefrontCorner[i][w]$, with time
$$t_w=(i-1)\tStepover + \frac{\wavefrontLength[i][w]}{\totalWavefrontLength[i]}\tStepover.$$
If $\texttt{Q}[w]$ is more than $\stepover'$ away from
$\spiral[\parentSpiralCorner[i-1][pw]]$, we redefine $\texttt{Q}[w]$ to be the point on the same path
with distance exactly $\stepover'$.
We mark the path from $\texttt{Q}[w]$ to the root of $\VD$.
See figure \ref{interpolationDetail}.

When we have done the marking for each $w$, we traverse wavefront $i$ once more.
For each wavefront corner $\wavefront[i][w]$,
we find the first marked point on the path
to the root. We let $\PPP[w]$ be that point and
$\texttt{T}[w]=\theta(\PPP[w])$ be its time. We have that $\texttt{T}[w] \geq t_w$, because
a later wavefront corner $\wavefront[i][w']$, $w'>w$, can mark more of the path from
$\wavefront[i][w]$ to the root. Therefore, we can have $\PPP[w]=\PPP[w+1]$ for some $w$.
Using this construction, there is exactly one distinct
$\PPP$-point on each path from a wavefront corner to the root.
Furthermore, we know that the distance from $\PPP[w]$ to the spiral corner
$\spiral[\parentSpiralCorner[i-1][\parentWavefrontCorner[i][w]]]$
is at most $\stepover'$.

The polyline defined by the points $\PPP[0],\PPP[1],\ldots$
is basically our interpolated spiral, but the points
have a tendency to have unnecessarily sharp corners if $\VD$ is relatively dense, which is
often the case for polygons $\PP$ occurring in real-world problems.
To avoid unnecessary sharp corners, we apply a method which we denote as the \emph{convexification},
see figure \ref{interpolationDetail}.
Let $\texttt{D}[w]=\sum_{v=0}^{w-1} \lVert\PPP[v]-\PPP[v+1]\rVert$ be the length of the polyline
until $\PPP[w]$ and consider the points $(\texttt{D}[w],\texttt{T}[w])$. We compute the upper
convex hull of these points, e.g. using the method of Graham and Yao \cite{graham1983}.
Let $\FF$ be the function whose graph
is the upper hull. By definition, we have that $\texttt{T}[w]\leq \FF(\texttt D[w])$
for each $w=0,\ldots$.
We now choose the corners of $\spiral$ in the following way:
For each wavefront corner $\wavefront[i][w]$ in order, we find the point $S$
on the path to the root with time
$\FF(\texttt D[w])$. If $S$ is more than $\stepover'$ away from the parent spiral corner
$\spiral[\parentSpiralCorner[i-1][\parentWavefrontCorner[i][w]]]$ (which will be the
parent of the spiral corner we are constructing), we
choose instead $S$ to be the point on the same path which is exactly $\stepover'$ away.
To avoid repetitions of the same point in $\spiral$, we add $S$ to the end of $\spiral$
if $S$ is different from the last point in $\spiral$.
Since we get the spiral corners by moving the $\PPP$-points closer to wavefront $i$,
we get exactly one distinct spiral corner on each path from a wavefront corner to the root.
When $\VD$ is sparse like in figure \ref{polylineSpiralFig}, the convexification makes no visible
difference between the $\PPP$-points and the final points in $\spiral$,
but when $\VD$ is dense as in figure \ref{interpolationDetail}, the effect is significant.

We also add one revolution around $\PP$ to the end of $\spiral$,
which is used to test that
the last interpolated revolution respects the stepover when the spiral is rounded later on.

The polyline spiral constructed as described clearly satisfies that the distance from
a point on one revolution to the neighboring revolutions is at most $\stepover'$.
Furthermore, each revolution is between two neighboring wavefronts
since all the corners of the interpolation between wavefronts $i$ and $i+1$ have times
in the interval $\left[i\tStepover,(i+1)\tStepover\right[$.
The wavefronts do not intersect as mentioned earlier, so different revolutions of
the polyline spiral do not either. It is also clear from the construction that
one revolution does not intersect itself.
Therefore, the polyline spiral has all the properties that we require of our spiral
except being $G^1$ continuous. How to obtain that is described in section
\ref{rounding}.

\subsection{Enriching the diagram}\label{enriching}

In this and the following section, we describe some modifications we make on the Voronoi
diagram of $\PP$ before doing anything else. The result is the diagram $\VD=\VD(\PP)$.
Long edges on $\PP$ lead to long
faces in the Voronoi diagram,
so that the wave is not moving towards the boundary $\partial\PP$
in a natural way, see figure \ref{wavefronts}.
In figure \ref{wavefrontsBad}, the wave starts on a long edge, and the first three
wavefronts are all degenerated polygons with only two corners, both on the edge.
Therefore, we need edges going directly to the boundary with a distance to each other
of at most $\stepover$,
so that each wavefront has corners on more edges than the previous one.
We have obtained this by traversing the Voronoi diagram and inspecting each pair of
consecutive leafs $l_1$ and $l_2$. Such a pair of nodes are
on the same or on two neighboring corners of $\PP$. Assume the latter, so that
there is a segment $S$ on $\partial\PP$ from $l_1$ to $l_2$ and a face $f$ of the Voronoi diagram
to the left of $S$.
Let $s=\point[l_2] - \point[l_1]$ be the vector from $l_1$ to $l_2$,
$d=\lVert s\rVert$ be the length of $S$ and
$m=\lceil d/\stepover\rceil$. We want to subdivide $f$ into $m$ faces.
Let
$p_i=\point[l_1] + s\cdot \frac i m$, $i=1,\ldots,m-1$, be interpolated points on $S$.
Let $h_i=\ora{p_i,p_i+\widehat s}$ be the half-line starting at $p_i$ with direction $\widehat s$,
where $\widehat s$ is the counterclockwise rotation of $s$.
For each $i=1,\ldots,m-1$, we find the first intersection point between $h_i$ and the Voronoi
diagram. Assume the intersection for some $i$ is a point $q$ on an edge $e$.
If the smallest angle between
$h_i$ and $e$ is larger than 50 degrees, we split $e$ into two edges by introducing a node at $q$
and add a segment from that node to a new node at $p_i$. If the smallest angle is less than 50 degrees,
the Voronoi diagram is moving fast enough towards the boundary so that the wavefronts will be fine
in that area without adding any additional edges.
Figure \ref{enrichedVD} shows the result of enriching the Voronoi diagram shown in
figure \ref{rawVD}.

\subsection{Removing double edges to concave corners}\label{removing}

A \emph{concave} corner of $\PP$ is a corner where the inner angle is more than 180 degrees.
Each concave corner $c$ on $\PP$ leads to a face in the Voronoi diagram of all the points
in $\PP$ being closer to $c$ than to anything else on the boundary of $\PP$.
Therefore, there are two edges $e_1$ and $e_2$ of the Voronoi diagram with an endpoint on $c$.
We have found that we get a better spiral if we remove these
edges and instead add an edge following the angle bisector of the edges, i.e., we follow the bisector
from $c$ and find the first intersection point $q$ with the Voronoi diagram and add an edge from
$q$ to $c$, see figure \ref{finalVD}.
The reason that this process improves the resulting spiral is that the wavefronts
will resemble $\PP$ more because they will have one corner on the
bisector edge corresponding to the corner $c$ on $\PP$.
We can only do this manipulation if the resulting
faces are also convex. That is checked easily by computing the new angles of the manipulated faces
and it seems to be the case almost always.

\subsection{Rounding the polyline spiral}\label{rounding}

\begin{figure*}
\centering
\includegraphics{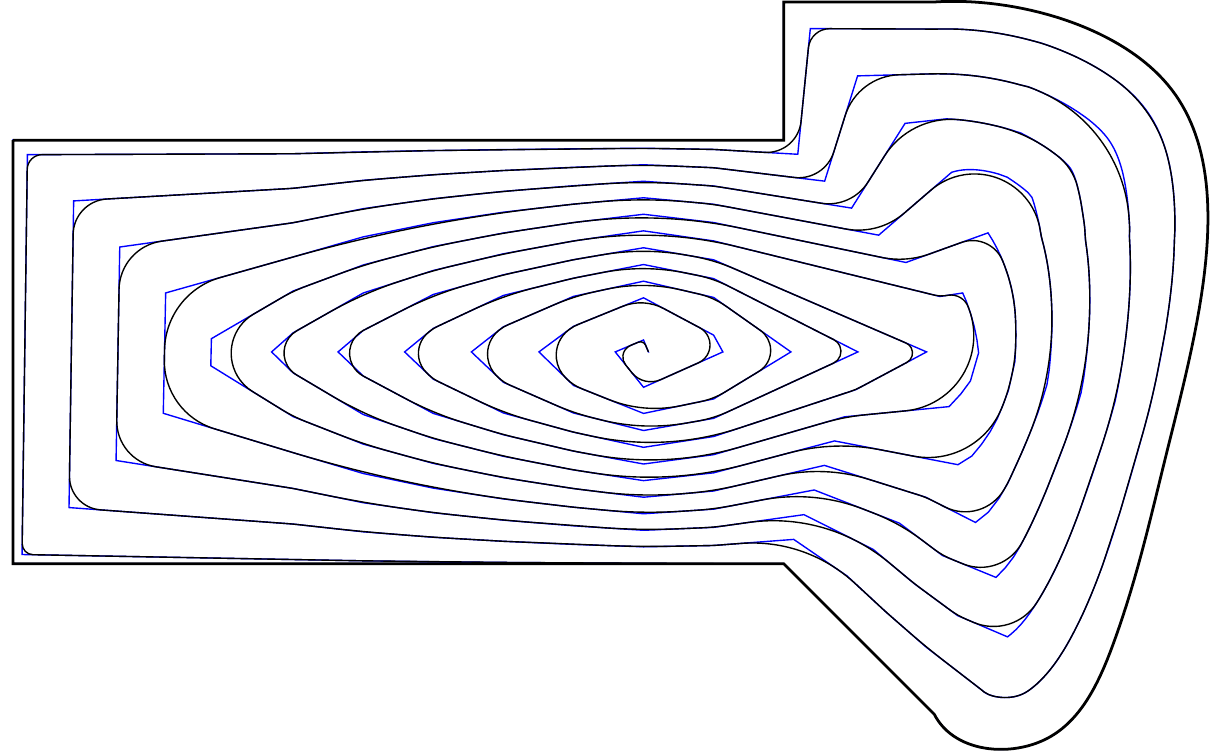}
\label{spiralRounding}
\caption{The spirals from figure \ref{firstSpirals} together. The polyline spiral is blue and the rounded
spiral is black.}
\end{figure*}

In this section we describe a possible way for smoothing the polyline spiral to get a
$G^1$ continuous spiral.
See figure \ref{spiralRounding} for a comparison between the rounded and unrounded spiral from figure
\ref{firstSpirals} using this method.
Note that some of the arcs of the rounded spiral rounds multiple corners
of the polyline spiral.

For each corner on the polyline spiral, we substitute a part of the spiral containing the corner with a
circular arc which is tangential to the polyline spiral in the endpoints.
That gives a spiral which is $G^1$ continuous, i.e., having no sharp corners. Each arc is either
clockwise or counterclockwise. For each index $i$, let
$s_i=\spiral[i]\spiral[i+1]$ be the segment from $\spiral[i]$ to $\spiral[i+1]$ and
$v_i=\spiral[i+1]-\spiral[i]$ be the vector from $\spiral[i]$ to $\spiral[i+1]$.
Each arc has the startpoint $p$ on some segment $s_a$ and the endpoint $q$
on another segment $s_b$, $a<b$, so that the arc \emph{substitutes} the part of the
polyline spiral from $p$ to $q$. We say that the arc \emph{rounds} the corners
$a+1$ to $b$.
We call the arc \emph{tangential} if it is counterclockwise and its center is on the intersection of the
half-lines $\ora{p,p+\widehat {v_s}}$ and $\ora{q,q+\widehat {v_r}}$ or it is
clockwise and its center is on the intersection of the half-lines
$\ora{p,p-\widehat {v_s}}$ and $\ora{q,q-\widehat {v_r}}$.

We store a pointer $\arc[i]$ to the arc that substitutes the corner $\spiral[i]$.
The same arc can substitute multiple consecutive corners, so that
$\arc[i]=\arc[i+1]=\ldots =\arc[i+k-1]$.
In that case, when $\arc[i-1]\neq\arc[i]\neq\arc[i+k]$,
the \emph{neighbors} of $\arc[i]$ are $\arc[i-1]$ and $\arc[i+k]$.
Two different arcs must substitute disjoint parts of the polyline spiral
for the rounded spiral to be well-defined. We subdivide each segment
$s_a$ at a point $p_a\in s_a$ such that an arc ending at
$s_a$ must have its endpoint at the segment $\spiral[a]p_a$ and an arc
beginning at $s_a$ must have its startpoint on the segment
$p_a\spiral[a+1]$. The point $p_a$ is chosen as a weighted average of $\spiral[a]$ and
$\spiral[a+1]$ so that the arc rounding the sharpest of the corners $\spiral[a]$ and $\spiral[a+1]$
gets most space. Let $\phi_a\in(-\pi,\pi]$ be the angle at corner $\spiral[a]$ of the polyline spiral.
We set $w_a=\frac{\pi-\phi_a}{2\pi-\phi_a-\phi_{a+1}}$ and choose $p_a$ as
$p_a=(1-w_a)\cdot \spiral[a] + w_a\cdot \spiral[a+1]$.

We keep a priority queue $Q$ \cite{cormen2009} of the arcs that can possibly be
enlarged.
After each enlargement of an arc, the resulting spiral respects the stepover $\stepover$
and no self-intersection has been introduced. We say that an arc with these properties is
\emph{usable}.
Initially, we let each corner be rounded by a
degenerated zero-radius arc, and $Q$ contains all these arcs.
Clearly, these initial zero-radias arcs are usable by definition.
We consider the front arc $A$ in $Q$ and try to find another usable arc $A'$
that substitutes a longer chain of the polyline spiral.
The new arc $A'$ normally has a larger radius than $A$.
If possible, we choose $A'$ so that it also
substitutes one or, preferably, two of the neighbors of $A$. Assume that we succeed
in making a usable arc $A'$ substituting both $A$ and its neighbors $B$ and $C$.
Now $A'$ rounds the union of the corners previously rounded by $A$, $B$, and $C$.
For all these corners, we update the $\arc$-pointers
and we remove $A$, $B$, and $C$ from $Q$.
We add $A'$ to $Q$. For every corner that $A'$ rounds,
we also add the arcs rounding the children and parents of that corner to $Q$,
since it is possible that these arcs can now be enlarged.
If no larger usable arc $A'$ is found, we just remove $A$ from $Q$.
The rounding process terminates when $Q$ is empty.

We test if an arc is usable by measuring the distance to the arcs
rounding the child and parent corners. That is easily done using elementary
geometric computations.

Here we describe how to find the largest usable arc rounding the corners
$a+1$ to $b$, $a<b$. We find the possible radii of tangential arcs beginning at
a point on $p_a\spiral[a+1]$ and ending at a point on $\spiral[b]p_b$ using elementary geometry.
There might be no such arcs, in which case we give up
finding an arc rounding these corners.
Otherwise, we have an interval $[r_{\min},r_{\max}]$ of radii of tangential arcs
going from $p_a\spiral[a+1]$ and ending on $\spiral[b]p_b$.
We check if the arc with radius $r_{\min}$ is usable.
If it is not, we give up. Otherwise, we check if
the arc with radius $r_{\max}$ is usable. If it is, we use it. Otherwise, we
make a binary search in the interval $[r_{\min},r_{\max}]$
after the usable arc with the largest radius. We stop when the binary search interval
has become sufficiently small, for instance $0.01\cdot \stepover$, and use the arc
with the smallest number in the interval as its radius.

The order of the arcs in $Q$ is established in the following way:
We have found that giving the arc $A$ the priority
$P(A)=r(A)/r(C_{\max})+1/s(A)$ gives good results, where
$r(A)$ is the radius of $A$, $s(A)$ is the size of the subtended angle from the center of $A$
in radians, and
$r(C_{\max})$ is the radius of the maximum circle contained in $\PP$.
The front arc in $Q$ is the one with smallest $P$-value.
We divide by $r(C_{\max})$ to make the rounding invariant when $\PP$ and $\stepover$ are scaled
by the same number.
$r(C_{\max})$ can be obtained from the Voronoi diagram of $\PP$, since the largest inscribed circle
has its center on a node in the diagram.
If $s(A)=0$, we set $P(A)=\infty$, since there is no corner to round.

We see that arcs with small radii or large subtended angles are chosen first for enlargement.
In the beginning when all the arcs in $Q$ have zero radius, the arcs in the sharpest corners are
chosen first because their degenerated arcs have bigger subtended angles -- even though the radius
is zero, we can still define the start and end angle of the arc according to the slope of the segments
meeting in the corner and thus define the subtended angle of the arc.

If two or three arcs are substituted by one larger arc each time we succeed in making a larger arc,
we are sure that
the rounding process does terminate, since the complexity of the spiral decreases.
However, it is often not possible to merge two or three arcs, but only to make a larger
arc rounding the same corners as an old one. The rounded spiral gets better, but we cannot
prove that the process terminates. In practice, we have seen fast termination in any tested example.
A possible remedy could be only to allow each arc to increase in size without rounding more corners
a fixed number of times.

\section{Computing a spiral in a pocket with an island}\label{pocketWithHole}

In many practical applications, the area to be machined is not simply-connected,
but has one or more ``islands'' that should not be machined. It might be because there are
physical holes in the part or areas of a thicker layer of material not to be machined in the same depth.
Therefore, assume that we are
given a polygon $\PP$ and a single polygonal island $\HH$ in the interior of $\PP$. In
section \ref{multipleHoles} we suggest a method to deal with multiple islands.
By $\PP\setminus\HH$, we denote the closed set of points which are in the interior or
on the boundary of $\PP$ but not in the interior of $\HH$.
We want to compute a spiral which
is contained in $\PP\setminus\HH$ such that the Hausdorff distance is at most
$\stepover$ between (i) two consecutive revolutions,
(ii) $\partial\HH$ and the first revolution, and (iii) $\partial\PP$ and the last revolution.
As before, $\stepover$ is the user-defined \emph{stepover}.
We also require
that the spiral is $G^1$ continuous and
has no self-intersections. See figure \ref{holeResult} for an example.

As in the case with a simply-connected pocket, we use a wave model to construct the spiral.
We imagine a wave that has exactly the shape of $\partial\HH$ at time $0$ and moves
towards $\partial\PP$, so that at the time $1$, it reaches $\partial\PP$ everywhere.
We explain how to define a polyline spiral. The spiral should be rounded by a method similar to the one
described in section \ref{rounding}. 

\subsection{The Voronoi diagram of a pocket with an island}

We use the Voronoi diagram of
the set of line segments of $\PP$ and $\HH$. As in the case with no islands, we modify the
diagram slightly. Let $\VD=\VD(\PP\setminus\HH)$ be the modified polygon. Like the true Voronoi diagram,
the modified diagram $\VD$ has the following properties:
\begin{itemize}
\setlength\itemsep{0em}
\item $\VD$ is a connected, plane graph contained in $\PP\setminus\HH$,
\item each leaf of $\VD$ is on the boundary of $\PP$ or $\HH$,
\item there is at least one leaf of $\VD$ on each corner of $\PP$ and $\HH$,
\item all the faces into which $\VD$ divides $\PP\setminus\HH$ are convex,
\item $\VD$ contains exactly one cycle,
the cycle is the locus of all points being equally close to $\partial\HH$ and $\partial\PP$, and
$\HH$ is contained in its interior.
\end{itemize}

\begin{figure*}
\centering
\subfigure[]{
\includegraphics[scale=0.98]{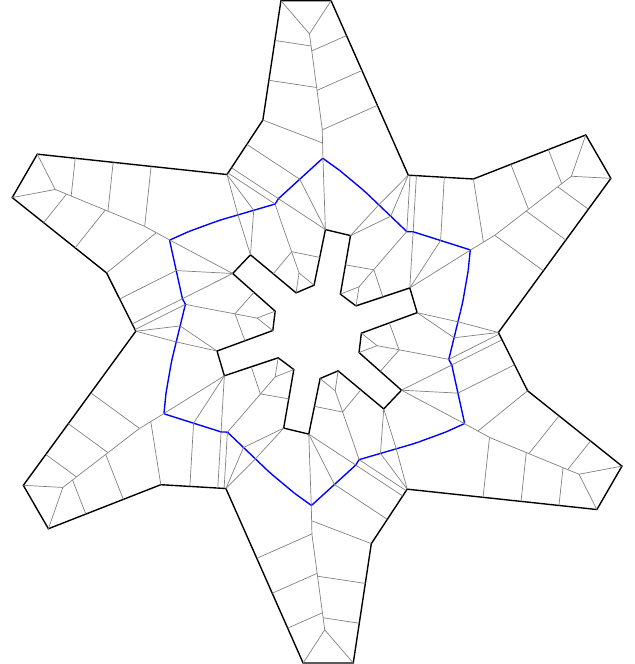}
\label{holeVD}
}\quad
\subfigure[]{
\includegraphics[scale=0.98]{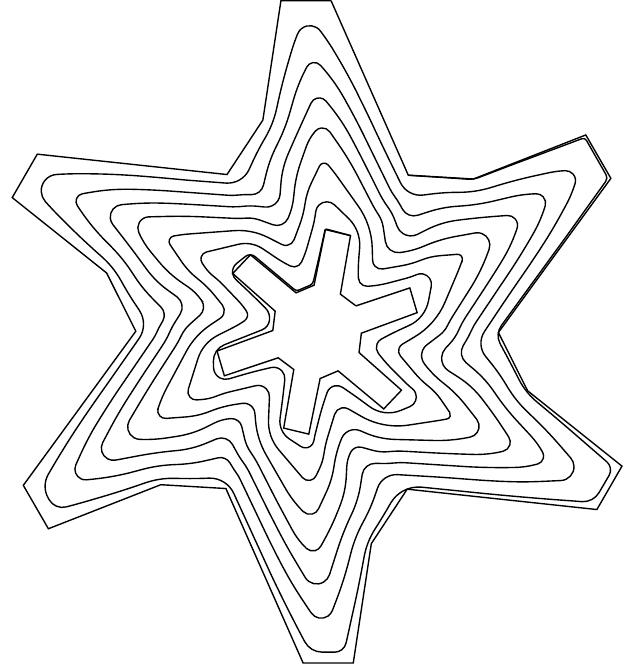}
\label{holeResult}
}
\caption{
\subref{holeVD} A polygon $\PP$ with an island $\HH$, both in black.
The diagram $\VD$ of $\PP\setminus\HH$ is drawn with the cycle $\CC$ in blue
and the other edges in gray.
\subref{holeResult} The resulting spiral.
}
\label{spiralWithHole}
\end{figure*}

The diagram $\VD$ is consisting of one cycle $\CC$, which we also denote by the \emph{central cycle},
and some trees growing out from $\CC$, see figure \ref{holeVD}.
Each of the trees grows either outwards and has all
its leafs on $\partial\PP$
or inwards and has all its leafs on $\partial\HH$.
The general idea is to use the method described in section \ref{withoutHoles}
to define wavefronts in each of these trees separately. Afterwards, we interpolate between
the neighboring wavefronts in each tree and connect the interpolated pieces to get one
contiguous spiral.

As in the case of a polygon without an island,
we enrich the Voronoi diagram by adding edges equidistantly along and perpendiculairly to long
edges on $\partial(\PP\setminus\HH)$ as described in section \ref{enriching}.
One small difference is that if the half-line $h_i$ intersects an edge $e$ at the point
$q$, and $e$ is an edge on the cycle
$\CC$, we do not split $e$ and do not introduce a new edge from $q$ to $p_i$.
There would be no advantage of adding that edge.
We also remove double edges to concave corners and add their bisector instead
as described in section \ref{removing}.

We want the trees to be \emph{symmetric} in the sense
that there is a tree $\PT_n$ with root $n\in\CC$ and leafs on $\partial\PP$ if and only if
there is a tree $\HT_n$ with root $n$ and leafs on $\partial\HH$. If for a node $n\in\CC$
we only have one of the trees, say $\HT_n$,
we add an edge from node $n$ to the closest point on $\partial\PP$ and let
$\PT_n$ be the tree consisting of that single edge.
It follows from the properties of the Voronoi diagram that the added edge does not intersect any of the
other edges.

We store $\CC$ as a vector $[n_0,\ldots,n_{c-1}]$ of the nodes on $\CC$
in counterclockwise order, such that there are trees $\HT_{n_i}$ and $\PT_{n_i}$ for each
$i=0,\ldots,c-1$.
A \emph{root node} is a node $n$ on $\CC$.
We let
$\TT_n=\PT_n\cup\HT_n$ be the union of the two trees rooted at node $n\in\CC$
and consider $\TT_n$ as a tree rooted at node $n$.

\subsection{Defining the movement of the wave}\label{defMovement}

On each tree $\TT_n$ we now define a wave model similar to the one described
in section \ref{waveModel}.
The wave starts at time $t=0$ on the leafs on $\partial\HH$ and moves through $\TT_n$ so that
it hits the leafs on $\partial\PP$ at time $t=1$.
Once the time and speed in the node $n$ is determined, the
times and speeds on the other nodes and edges in $\TT_n$ are computed analogously to
algorithm \ref{setTimesAndSpeeds}. One difference is that we compute decreasing times
for the tree $\HH_n$. A way to do so is to set $\ttimeO[n]=1-\ttimeO[n]$,
compute the inverse times using algorithm \ref{setTimesAndSpeeds}, and
afterwards inverse each of the computed times $t$ by setting $t=1-t$.

Let the \emph{preferred time} of a node $n$ be
$t_n=\frac{\holeHeight[n]}{\boundaryHeight[n] + \holeHeight[n]}$,
where $\boundaryHeight[n]$ is the length of the longest path to a leaf in $\PT_n$ and
$\holeHeight[n]$ is the length of the longest path to a leaf in $\HT_n$.
Similarly, if the time $\ttimeO[n]$ has already been defined,
we let the \emph{preferred speed} of $n$ be
$v_n=\max\left\{\frac{\holeHeight[n]}{\ttimeO[n]},
\frac{\boundaryHeight[n]}{1-\ttimeO[n]}\right\}$.

\begin{figure*}
\centering
\subfigure[]{
\includegraphics[scale=0.68]{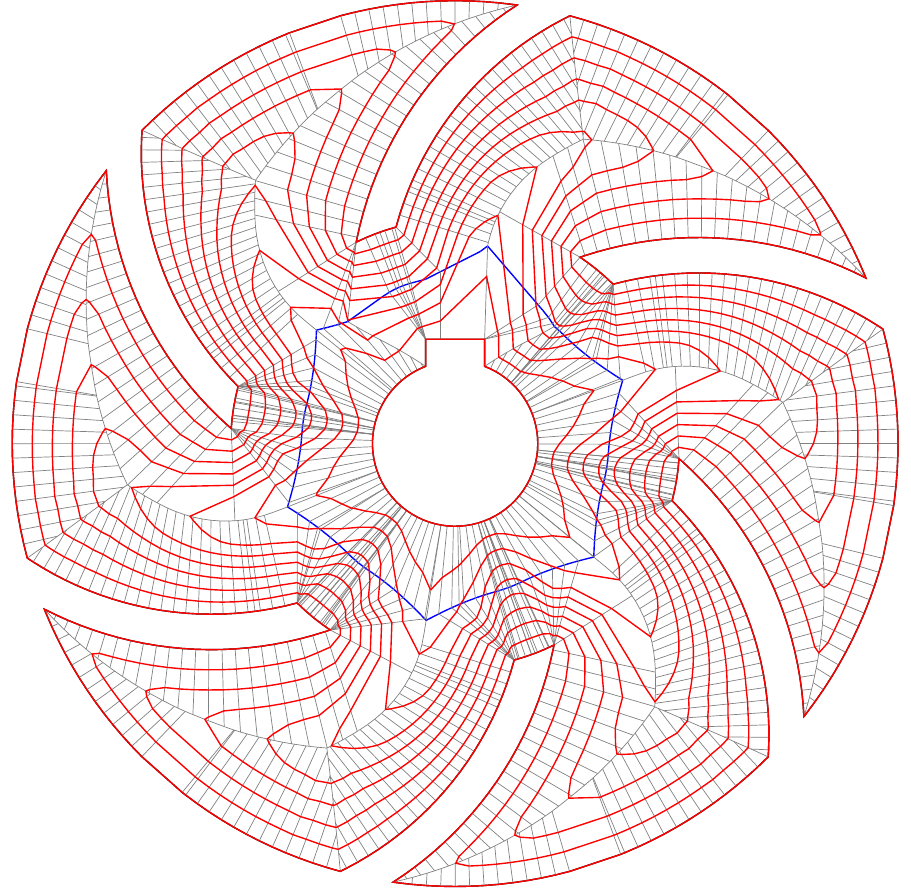}
\label{goodWavefronts}
}\quad
\subfigure[]{
\includegraphics[scale=0.68]{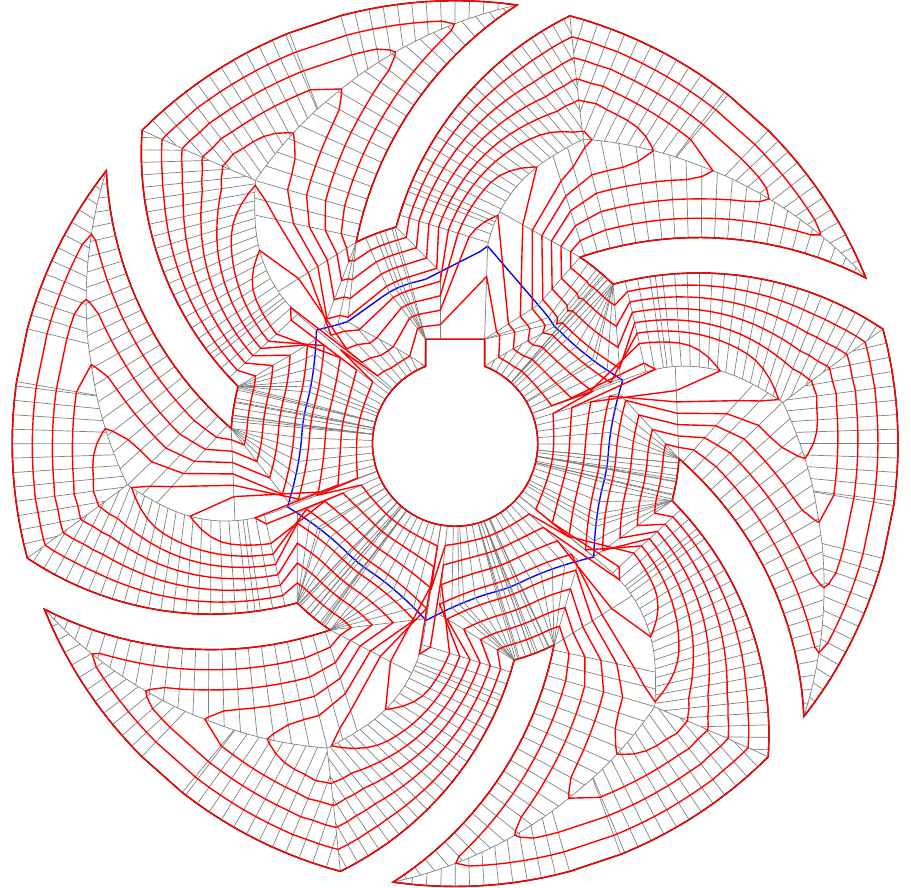}
\label{badWavefronts}
}
\caption{
\subref{holeVD} Wavefronts in red when the times and speeds of the wave in the root nodes
have been defined as described in \ref{appB}. $\CC$ is in blue and the rest of $\VD$ is in gray.
\subref{badWavefronts} Wavefronts when the preferred times and speeds have been used
to define the movement of the wave.
}
\label{holeWavefronts}
\end{figure*}

A naive method to define the times and speeds of a root $n$ is to set
$\ttimeO[n]=t_n$ and $\speedO[n]=v_n$.\comm{\begin{equation*}
\begin{split}
&\ttimeO[n]=t_n,\\
&\speedO[n]=v_n.
\end{split}
\end{equation*}} That will
minimize the number of revolutions and give the most equidistant
wavefronts on each tree $\TT_n$. However, the abrupt changes in time and speed along the central cycle
$\CC$ results in a spiral which curves a lot. Instead, we might smooth the times and
speeds around $\CC$. See figure \ref{holeWavefronts}
for a comparison of the wavefronts with and without smoothing.
There may be many ways of smoothing the values. After much experimentation, we
have found the method described in \ref{appB} to give good results.
Notice that the wavefronts in figure \ref{badWavefronts} intersect each other, and
that will indeed also be the case for some problem instances after using the
smoothing method described in \ref{appB}.
However, in the end of section \ref{interpolWithIsland} we describe how to ensure that the
polyline spiral will not have any self-intersections.

\subsection{Creating wavefronts}

For a given root node $n\in\CC$, we want at least
$s_{\HH_n}=\frac{\holeHeight[n]}{\stepover'}$
revolutions of the spiral in the tree $\HT_n$ in order to respect the stepover $\stepover'=0.95\cdot\stepover$.
Similarly, we want $s_{\PP_n}=\frac{\boundaryHeight[n]}{\stepover'}$ revolutions in
$\PT_n$. Therefore, the time between two revolutions should be
at most $\tStepover_n=\min\{\frac{\ttimeO[n]}{s_{\HH_n}},\frac{1-\ttimeO[n]}{s_{\PP_n}}\}$.
Hence, we let $\tStepover'=\min_{n\in\CC}\{\tStepover_n\}$ be the minimum over all such values.
We let the number of revolutions be $r=\lceil 1/\tStepover'\rceil$ and set
$\tStepover=1/r$.

Each tree $\TT_n$ contains a contiguous subset of the
corners of any wavefront $i$. The corners of the subset are
points on $\HT_n$ if and only if $t\leq \ttimeO[n]$, otherwise they are on $\PT_n$.
Let $r_{n}=\lfloor\frac{\ttimeO[n]}{\tStepover}\rfloor$.
The wavefronts $i=0,\ldots,r_{n}$ are on
$\HT_n$, while wavefronts $i=r_{n}+1,\ldots, r$ are on
$\PT_n$.
As explained in section \ref{defMovement}, we do not use the same time $\ttimeO[n]$ for every
node $n$ on $\CC$. Therefore, we may get wavefronts crossing $\CC$, as
is seen in figure \ref{holeWavefronts}.

We suggest to store the wavefront corners in a two-dimensional array for each of the trees
$\HT_n$ and $\PT_n$.
The wavefront corners on $\HT_n$ of one wavefront are stored in an array $\holeWavefront[n][j]$,
where index $j$ corresponds to wavefront $i=r_{n}-j+1$.
In $\PT_n$, the array $\boundaryWavefront[n][j]$ stores corners of
wavefront $i=j+r_{n}$.
Hence, the corners of each wavefront is stored locally in each tree $\TT_n$.

In $\HT_n$, the parents of the corners of wavefront $i=0,\ldots,r_{n}-1$
are corners of wavefront $i+1$. In $\PT_n$, the parents of the corners of wavefront
$i=r_{n}+2,\ldots, r$ are corners of wavefront $i-1$.
Therefore, all parents are on the
wavefront one step closer to the root $n$.
In both $\HT_n$ and $\PT_n$, we introduce fake wavefront corners
at node $n$ stored in the arrays $\holeWavefront[n][0]$ and $\boundaryWavefront[n][0]$,
respectively, which are the parents of the corners in
the arrays $\holeWavefront[n][1]$ and $\boundaryWavefront[n][1]$.
Thus, these fake corners are not corners on wavefront $i$ for any $i=0,\ldots, r$, but are merely
made to complete the tree of parent pointers between corners of neighboring wavefronts.

We also need an array $\wavefrontLength$ containing \emph{global} information about the length
of each wavefront crossing all the trees $\{\TT_{n}\}_n$ in order to do interpolation between
the wavefronts later. We have
$\wavefrontLength[i][0]=0$ for every wavefront $i$.
If $c_m$ and $c_{m+1}$ are the $m$'th and $(m+1)$'st corners on
wavefront $i$, respectively, we have
$\wavefrontLength[i][m+1]=\wavefrontLength[i][m] +
\lVert c_{m+1} - c_m \rVert$. Notice that $c_m$ and $c_{m+1}$ can be corners
in different, however neighbouring trees $\TT_n$ and $\TT_{n'}$ and hence stored in
different arrays.
$\totalWavefrontLength[i]$ stores the total length of wavefront $i$.

\subsection{Interpolating between wavefronts}\label{interpolWithIsland}

We interpolate between two wavefronts $i-1$ and $i$ in each tree $\TT_{n}$ separately,
but using the same technique as in section \ref{interpolating}. If
$i\leq r_{n}$, we interpolate between the
wavefront fragments stored in
$\holeWavefront[n][j]$ and $\holeWavefront[n][j+1]$,
where $j=r_{n}-i+1$, using the values of the length of wavefront $i-1$ stored in
$\wavefrontLength[i-1]$.
If $i>r_{n}+1$, we interpolate between
$\boundaryWavefront[n][j-1]$ and $\boundaryWavefront[n][j]$,
where $j=i-r_{n}$, using the values stored in $\wavefrontLength[i]$.
A special case occurs when
$i=r_{n}+1$, i.e.,
when we are interpolating between the first wavefront on each side of the root node $n$.
In that case, let
$$t=(i-1)\tStepover + \frac{\wavefrontLength[i][m]}{\totalWavefrontLength[i]}\tStepover,$$
when $\boundaryWavefront[n][1][0]$ is the $m$'th corner on wavefront $i$.
If $t\leq \ttimeO[n]$, we interpolate between $\holeWavefront[n][0]$ and $\holeWavefront[n][1]$.
Otherwise, we interpolate on the other side of $\CC$, that is,
between $\boundaryWavefront[n][0]$ and $\boundaryWavefront[n][1]$.
The convexification process described in section \ref{interpolating} can be used in each
tree $\TT_n$ separately.

We may store the interpolated spiral in an one-dimensional array $\spiral$.
Before we add the first interpolated revolution to $\spiral$, i.e., the one between wavefront $0$ and $1$,
we add wavefront $0$ to $\spiral$, that is, all the corners on $\partial\HH$. Likewise,
after the final revolution between wavefronts $r-1$ and $r$, we add wavefront $r$, which is
all the corners on $\partial\PP$. These are used to ensure that the distance from the
first and last revolution to $\partial\HH$ and $\partial\PP$, respectively,
does respect the stepover when rounding the spiral.

\begin{figure*}
\centering
\includegraphics{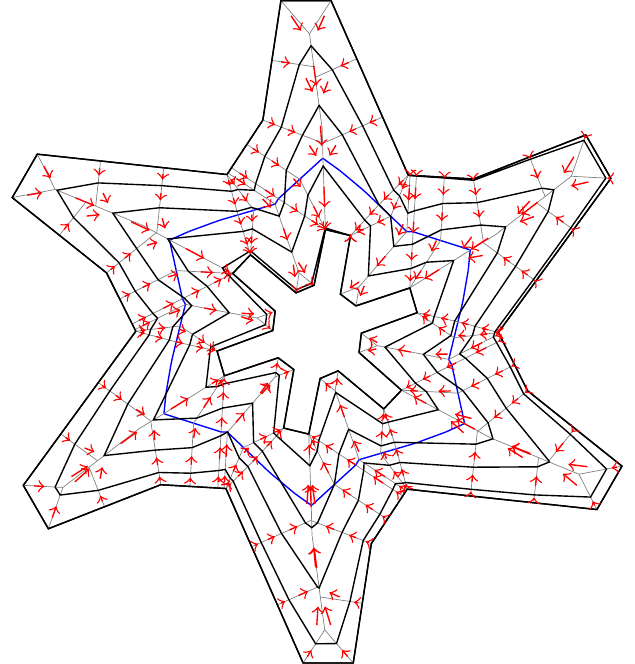}
\caption{
A polyline spiral in black and red arrows from each corner $\spiral[s]$ to its parents.
The cycle $\CC$ is in blue and the diagram $\VD$ is in gray.
}
\label{holePolyline}
\end{figure*}

For every corner $\spiral[s]$ we have a pointer $\parent[s]$ such that $\spiral[\parent[s]]$ is
the first spiral corner we meet when traveling from $\spiral[s]$ to the root $n$ of the tree $\TT_n$
containing $\spiral[s]$.
The parents are not defined for the spiral corners closest to the root node $n$.
Therefore, we make every spiral corner between $\holeWavefront[n][0]$ and
$\holeWavefront[n][1]$
the parent of every spiral corner between $\boundaryWavefront[n][0]$ and $\boundaryWavefront[n][1]$
to make parent dependencies across $\CC$.
We want all the parent pointers to be towards the island $\HH$. Therefore, we reverse
all the pointers between pairs of corners in each tree $\HT_n$. Now, the parent pointers
are defined for all spiral corners except for the ones on $\partial\HH$.
See figure \ref{holePolyline} for an example of a polyline spiral around an island and red arrows
indicating the parent pointers. Notice that a corner can have multiple parents, but the parents
are consecutive and can thus be stored using two indices.

\begin{figure*}
\centering
\subfigure[]{
\includegraphics[scale=0.8]{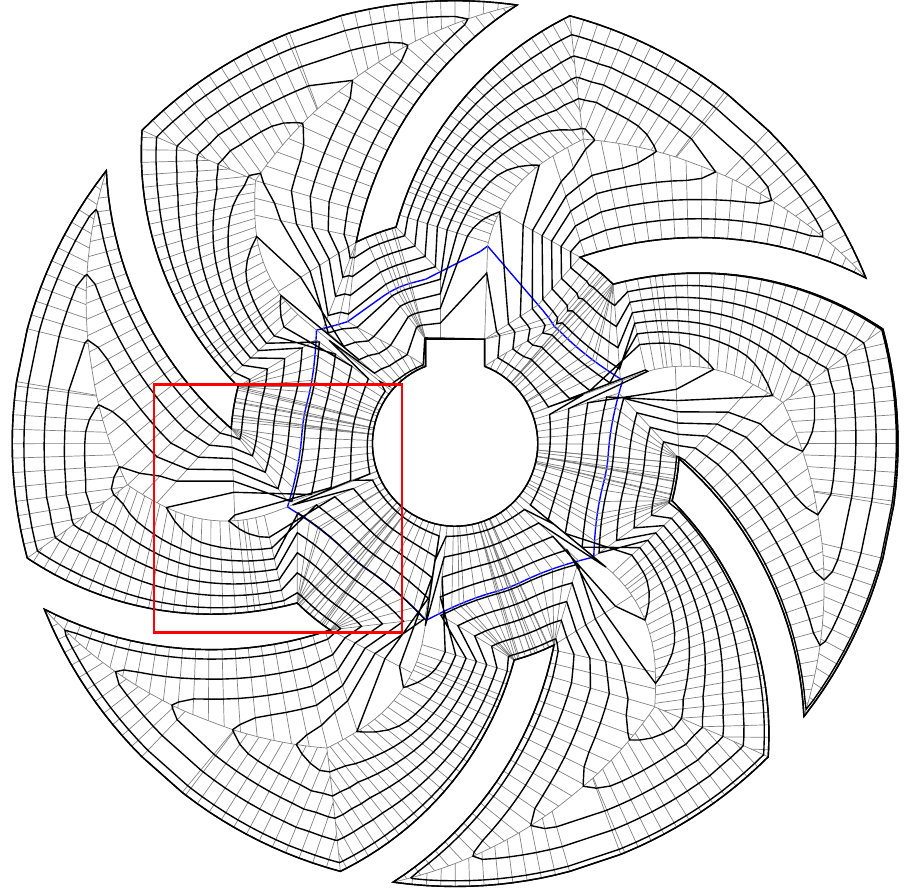}
\label{crossing}
}\\
\subfigure[]{
\includegraphics[scale=0.72]{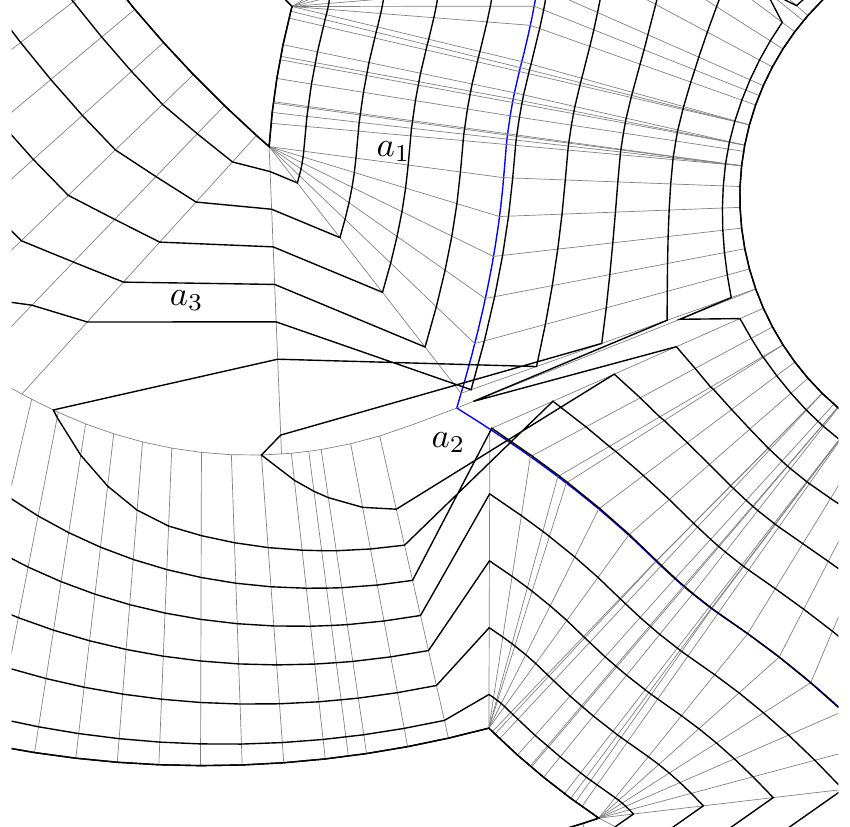}
\label{crossingZoom}
}\quad
\subfigure[]{
\includegraphics[scale=0.72]{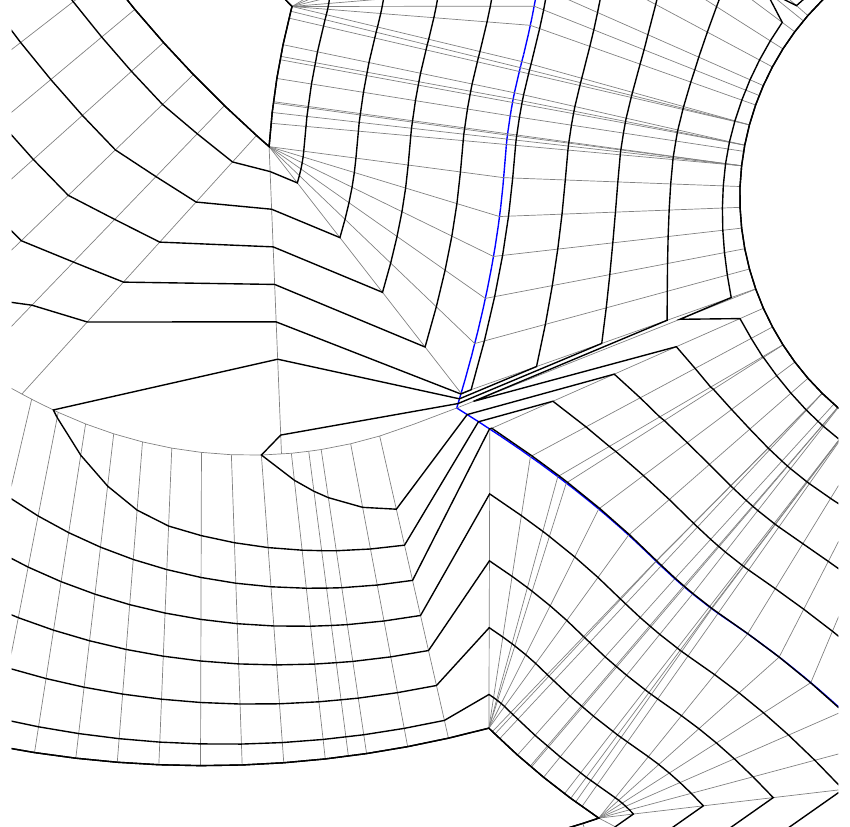}
\label{noCrossingZoom}
}
\caption{
The same pocket and island as in figure \ref{holeWavefronts}.
The cycle $\CC$ is blue, the other edges of $\VD$ are gray.
The interpolated spiral is black, and we have not introduced extra spiral corners on
$\CC$ to avoid self-intersections.
In order to emphasize the intersection problems that can arise,
we have not used the method described in \ref{appB},
but merely used the preferred time and speed of each node (section \ref{defMovement}).
\subref{crossingZoom} is a close-up of \subref{crossing}
of the area in the red rectangle. In \subref{noCrossingZoom}, we have introduced new corners on
$\CC$ when the spiral jumps from one side of $\CC$ to the other when the union of the two
faces on each side of $\CC$ is not convex.
}
\label{repairCrossings}
\end{figure*}

In section \ref{interpolating}, we stated that since there are no intersections between different
wavefronts,
the polyline spiral has no self-intersections when there is no island in $\PP$.
The wavefronts do not intersect in that
case due to the convexity of the faces into which the diagram $\VD(\PP)$ subdivides $\PP$.
When there is an island $\HH$ in $\PP$,
there are two kinds of faces into which $\VD(\PP\setminus\HH)$ subdivides $\PP\setminus\HH$,
see figure \ref{repairCrossings}. Some faces, like $a_3$ in the figure, are bounded
by edges of one tree $\TT_n$ while some are between two trees,
like $a_1$ and $a_2$. The latter kind is bounded by edges of two
neighboring
trees $\TT_n$ and $\TT_m$ and an edge $e$ from $n$ to $m$ on $\CC$, where $n$ and $m$ are neighboring nodes on $\CC$.
The first kind of faces is similar to the faces in section \ref{withoutHoles},
so here we do not worry about
self-intersections of the spiral. The second, however, can lead to wavefronts crossing each other and
therefore also a self-intersecting spiral as is the case in the figure when the spiral jumps over $\CC$ and
crosses $a_2$. If the union of
the faces on each side of $e$ is convex,
like $a_1$ and its neighboring face on the other side of $\CC$, there is no problem.
It can easily be tested if the union of the faces is convex by considering the angles of the union
at nodes $n$ and $m$. If it is not, we introduce a new corner
on the edge on $\CC$ whenever the spiral jumps from one side of $\CC$
to the other. The new corner is an interpolation of nodes $n$ and $m$
using the time of the spiral in the last corner in tree $\TT_n$.
Figure \ref{noCrossingZoom} shows the result of introducing
the extra corners. Our experience is that these intersection problems occur very rarely
when the method described in \ref{appB} has been applied.

\section{The skeleton method for a spiral in a pocket without an island}\label{skeletonMethod}

\begin{figure*}
\centering
\subfigure[]{
\includegraphics[scale=0.99]{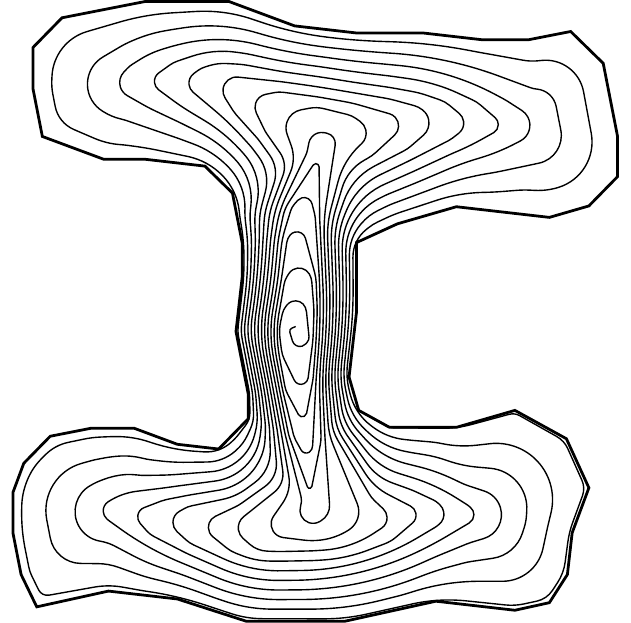}
\label{spiralNoHole}
}\quad
\subfigure[]{
\includegraphics[scale=0.99]{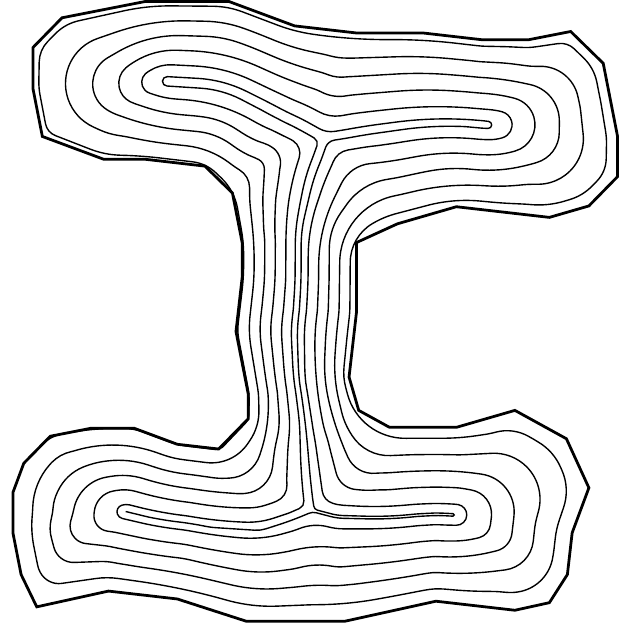}
\label{spiralSkeletonHole}
}
\caption{
Comparison of the basic spiral method from section \ref{withoutHoles}, figure \subref{spiralNoHole},
with the improved skeleton method from section \ref{skeletonMethod}, figure \subref{spiralSkeletonHole}.
Note that the spiral obtained from the skeleton method is significantly shorter and that the distance between
neighboring revolutions is varying much less than when using the basic method.
}
\label{skeletonHole}
\end{figure*}

The method from section \ref{withoutHoles} is mainly applicable
if the polygon $\PP$ is not too far from being a circle. If $\PP$ is very elongated or branched,
the distance between neighboring revolutions will often be much less
than the maximum stepover. Therefore, the toolpath will be unnecessarily long and the cutting width will
vary a lot. That leads to long machining time and an uneven finish of the part.
See figure \ref{skeletonHole} for an example.
In such cases, we construct a \emph{skeleton} in $\PP$,
which is an island $\HH$ with zero area.
We then use the method from section \ref{pocketWithHole} to make a spiral from the island to the boundary.
It does not matter for the construction of the spiral that the island $\HH$ has zero area.
See figure \ref{skeletonHole} for a comparison between the basic method from
section \ref{withoutHoles} with the skeleton method described here when applied to the same
polygon.

\subsection{Constructing the skeleton in a polygon $\PP$}

\begin{figure*}
\centering
\subfigure[]{
\includegraphics[scale=0.99]{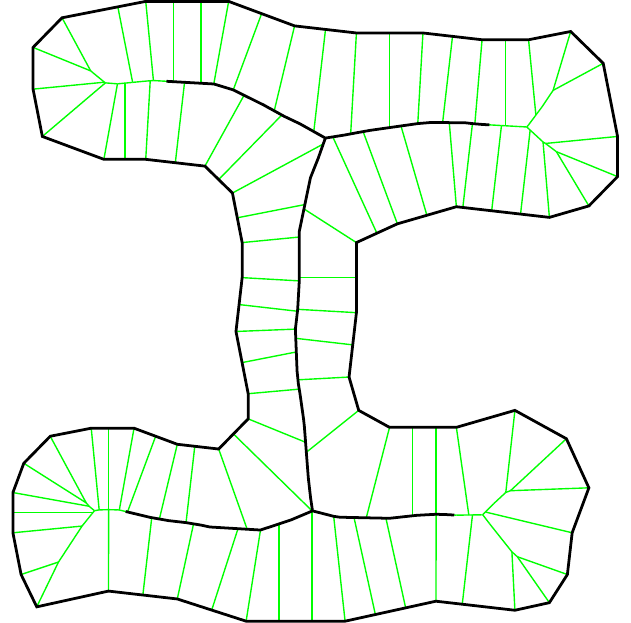}
\label{spiralHoleConstruction}
}\quad
\subfigure[]{
\includegraphics[scale=0.99]{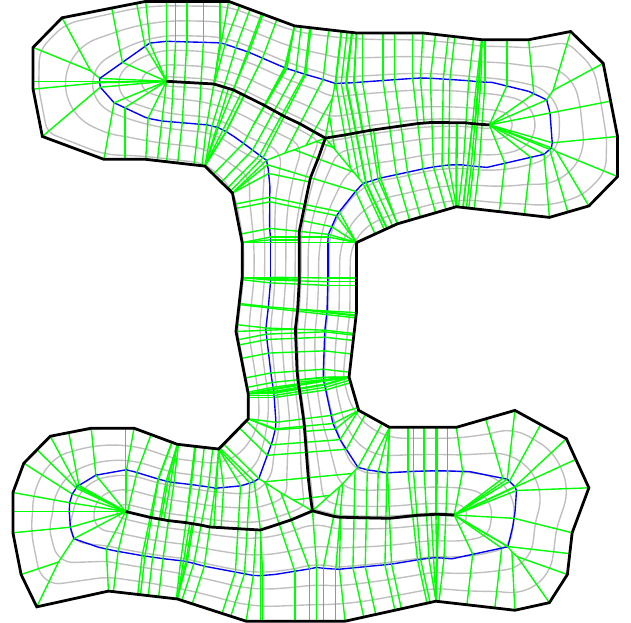}
\label{spiralSkeletonVD}
}
\caption{
\subref{spiralHoleConstruction} The diagram $\VD(\PP)$ of the polygon $\PP$ from figure \ref{skeletonHole} in green,
where the edges chosen for the skeleton are black. \subref{spiralSkeletonVD} The diagram
$\VD(\PP\setminus\HH)$ of
$\PP$ with the skeleton considered an island $\HH$. The cycle $\CC$ is blue and the remaining edges are green.
The resulting spiral from figure \ref{spiralSkeletonHole} is included in gray.
}
\label{skeletonConstruction}
\end{figure*}

We choose the skeleton as a connected subset of the edges of the diagram $\VD=\VD(\PP)$.
We traverse $\VD$ once starting at the root and decide for each edge whether to include it in the
skeleton. If an edge from node $n$ to $m$ is not included, we don't include anything from the sub-tree
$\VD[m]$.
For any node $n$, let $d(n)$ be the length of the shortest path from $n$ to a leaf in $\VD[n]$, and
let $D=\max_{n\in\VD} d(n)$. We have found that the following criteria for including
an edge $e$ from node $n$ to $m$ gives good results. We require all the criteria to be satisfied.
\begin{enumerate}
\setlength\itemsep{0em}
\item \label{case0} The longest path from $n$ to a leaf in $\VD[n]$ goes
through $m$ or $\ell[e]+\height[m]\geq 1.5\cdot D$, where $\ell[e]$ is the length of edge $e$.

\item \label{case1} The length of the spanned boundary (defined in \ref{appB})
of $m$ is larger than $2\cdot D$.

\item \label{case2} $\height[m]\geq D$.
\end{enumerate}

Criterion \ref{case0} is to avoid getting a skeleton that branches into many short paths. Therefore,
we only make a branch which is not following the longest path from $n$ if it seems to
become at least $0.5\cdot D$ long
(taking criterion \ref{case2} into account).
When criterion \ref{case1} fails,
it seems to be a good indicator that an edge is not a significant, central edge
in $\VD$, but merely one going straight to the boundary.
Criterion \ref{case2} ensures that we do not get too close to the boundary. If we did, we would get
very short distances between the neighboring revolutions there. If
criterion \ref{case2} is the only failing criterion, we find the point $p$ on $e$ such that
$\height[m]+\lVert p-\point[m]\rVert = D$ and include the edge from node $n$ to $p$
in the skeleton. See figure \ref{skeletonConstruction} for the skeleton constructed
given the polygon from figure \ref{skeletonHole}.

If the polygon is close to being a circle, the method described here
results in a very small skeleton, and we get a better spiral using the
basic method from section \ref{withoutHoles} in that case.
This can be tested automatically by falling back to the basic method if
the circumference of the skeleton is less than, say, 5\% of the circumference of $\PP$.

\section{Computing a spiral in a pocket with multiple islands}\label{multipleHoles}

\begin{figure*}
\centering
\subfigure[]{
\includegraphics[scale=0.96]{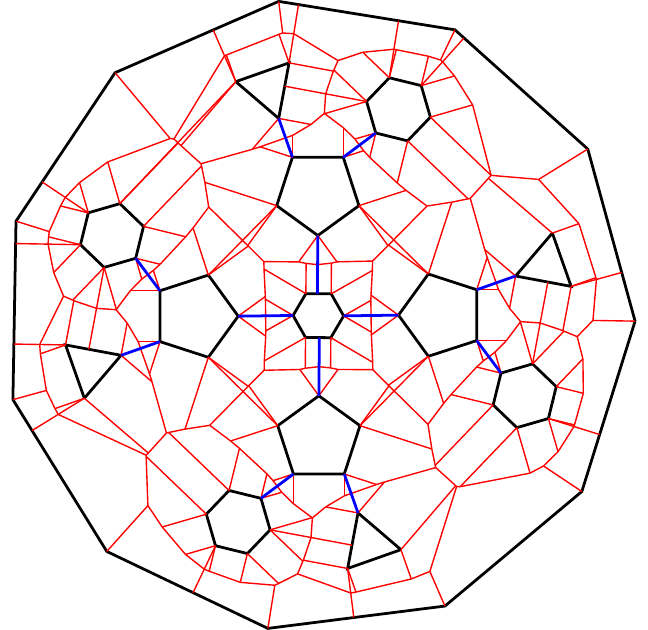}
\label{manyHolesConnection}
}\quad
\subfigure[]{
\includegraphics[scale=0.96]{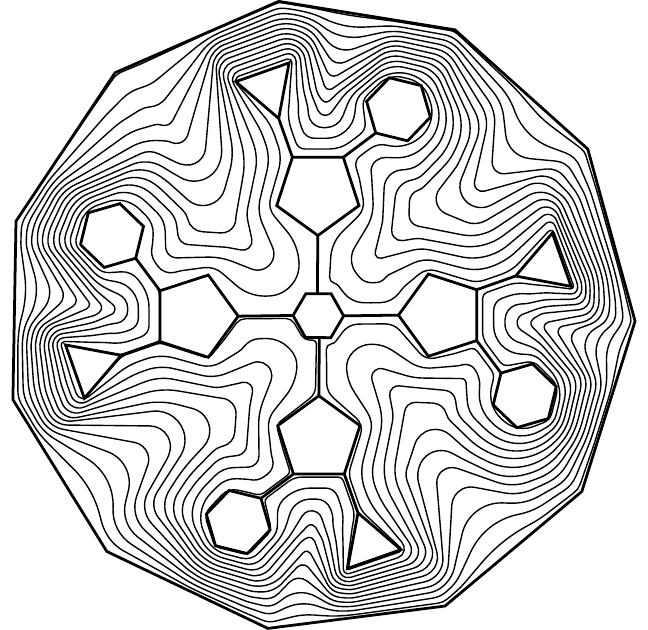}
\label{manyHolesResult}
}
\caption{
\subref{manyHolesConnection} A polygon with 13 islands. The bridges chosen among the Voronoi
edges to connect the islands are blue. The remaining Voronoi edges are red.
\subref{manyHolesResult} The resulting spiral around the islands.
}
\label{manyHoles}
\end{figure*}

The method from section \ref{pocketWithHole} works only for polygons with a single island.
If there are many islands $\HH_0,\ldots,\HH_{h-1}$ in a polygon $\PP$,
we may connect them with bridges in a tree structure to form one
big connected island, see figure \ref{manyHoles}.
The basic idea of reducing the number of islands by connecting them is also used
by Chuang and Yang \cite{chuang2007} and Held and Spielberger \cite{held2014}.
Our method is given in algorithm \ref{makeBridges}, and it is a variation of the minimum spanning
tree algorithm of Dijkstra \cite{dijkstra1959}.
We choose the bridges as edges in the Voronoi diagram $V$ of the area
$\PP\setminus \bigcup_{i=0}^{h-1}\HH_i$.
The algorithm creates an array $\bridges$ of the edges to use as bridges.
We keep a growing set $s$ of the nodes of the Voronoi diagram that we have
connected by bridges so far. Here, we have represented $s$ as a bit-vector.
We first find one central node $n_0$ and $s$ is only containing $n_0$ in the beginning.
We use Dijkstra's algorithm \cite{dijkstra1959} in the loop beginning at line \ref{MBwhile1}
to make all shortest paths from nodes in $s$. We make a relaxation of the distances in line \ref{MBrelaxation}. When we reach
an island $\HH_i$ whose corners are not in $s$,
we use the shortest path to that island as a bridge and add the nodes on the shortest path
and the corners on $\HH_i$ to $s$. The use of the distance vector $d$ makes the algorithm
prefer to build bridges from the vertices that has been in $s$ for the longest time.
That makes the bridges grow from the center node $n_0$ out in every direction.
If we did not choose the bridges in this careful way, the resulting connected island
would possibly make unnecessarily long dead ends that would require many
revolutions to fill out by the spiral.

\texttt{VRONI} by Held \cite{held2001}
has the possibility to add edges to the Voronoi diagram between
neighboring objects in the input where the distance between the objects is shortest,
even though these are not
genuine Voronoi edges. It is an advantage also to consider these edges when choosing
the bridges between the islands,
since they are often the best bridges. For instance, all the bridges chosen in the
example of figure \ref{manyHoles} are of that kind.

\begin{algorithm}[h]
\scriptsize
\LinesNumbered
\DontPrintSemicolon
\SetArgSty{}
\SetKwIF{If}{ElseIf}{Else}{if}{}{else if}{else}{end if}
\SetKwFor{Foreach}{for each}{}{end for}
\SetKwInput{Effect}{Effect}
Compute the Voronoi diagram $V$ of the area inside $\PP$ but outside every $\HH_i$.\;
Find the set of all nodes $N$ of $V$ which are either a corner on an island or
equally close to two islands but farther from
$\partial\PP$.\;
Let $n_0$ be the node in $N$ closest to the root of $\VD(\PP)$ defined as in section \ref{startingPoint}.\;
Let $s[n]\gets\texttt{FALSE}$ for each node $n\in V$.\;
Let $d[n]\gets\infty$ for each node $n\in V$.\;
$s[n_0]\gets\texttt{TRUE}$; $d[n_0]\gets 0$\;
$\bridges\gets [\; ]$\;
\While {$s$ does not include every island} { \nllabel{MBwhile0}
  For each node $n$, if $s[n]=\texttt{FALSE}$, set $d[n]=\infty$.\;
  Let $p[n]=\texttt{NULL}$ for each node $n$.\;
  Let $Q$ be a priority queue of the nodes $n$ of $V$ sorted after the values $d[n]$.\;
  \While {$Q\neq \emptyset$} { \nllabel{MBwhile1}
    Let $n$ be the front node in $Q$ and remove $n$ from $Q$.\;
    \If {$n$ is a corner on an island $\HH_i$ and $s[n]=\texttt{FALSE}$} {
      For each node $n'$ on the path $n,p[n],p[p[n]],\ldots$, set $s[n']\gets\texttt{TRUE}$.\;
      Add each edge on the path $n,p[n],p[p[n]],\ldots$ to $\bridges$.\;
      Let $n_f$ be the node on $\HH_i$ farthest from $n$ and let $a\gets\lVert n_f-n\rVert$.\;
      For each node $n'$ on $\HH_i$, set $d[n']\gets d[n]+a$ and $s[n']\gets \texttt{TRUE}$.\;
      Go to line \ref{MBwhile0}.\;
    }
    For each edge going out of $n$ to a node $n'\in V$,
    if $s[n']=\texttt{FALSE}$ and $d[n]+\lVert n-n'\rVert<d[n']$,
    set $d[n']\gets d[n]+\lVert n-n'\rVert$, update $Q$ accordingly, and set $p[n']\gets n$.\;  \nllabel{MBrelaxation}
  }
}
\caption{$\makeBridges(\PP,\HH_0,\ldots,\HH_{h-1})$}
\label{makeBridges}
\end{algorithm}

\section{Conclusion}\label{conclusion}
We have described methods for the computation of spiral toolpaths suitable for many
shapes of pockets for which no previously described algorithms yield equally good results.
Our main contribution is the new possibility of making a spiral morphing an island in a pocket
to the boundary of the pocket.

Our developed spiral algorithms only work for polygonal input. We believe that it should be possible to
generalize to input consisting of line segments and circular arcs, as done by
Held and Spielberger \cite{held2009}, using \texttt{ArcVRONI} by Held and Huber \cite{held2009topology}
to compute the Voronoi diagrams for such input.

Held and Spielberger \cite{held2014} developed methods to subdivide a pocket with arbitrarily many
islands into simply-connected sub-pockets, each of which are suitable for basic spirals as the ones
described in section \ref{withoutHoles}. Since we have the possibility to make spirals around islands,
we never have to partition the input into separate areas, but in some cases, for instance if the
pocket has a long ``arm'' requiring a lot of revolutions, it might be useful to machine
different areas of the input independently. Another possibility is to combine the method for
machining around multiple islands with the skeleton method.
Given these new possibilities, our problem is quite different from the one
discussed by Held and Spielberger. It would be interesting to investigate if and how one can
make an automatic partitioning that leads to a more efficient toolpath using our new kinds of spirals.

\section*{Acknowledgements}

I would like to thank my colleague at Autodesk,
Niels Woo-Sang Kj\ae rsgaard, for numerous helpful discussions while I did
the development of the spiral algorithms and for many useful comments on this paper.
My thanks also go to Autodesk in general for letting me do this research and publish the
result.

\bibliography{proj}{}

\begin{thebibliography}{10}

\bibitem{banerjee2012}
A.~Banerjee, H.-Y. Feng, and E.V. Bordatchev.
\newblock Process planning for floor machining of 2$\frac 12${D} pockets based
  on a morphed spiral tool path pattern.
\newblock {\em Computers \& Industrial Engineering}, 63(4):971--979, 2012.

\bibitem{bieterman2003}
M.B. Bieterman and D.R. Sandstrom.
\newblock A curvilinear tool-path method for pocket machining.
\newblock {\em Journal of Manufacturing Science and Engineering, Transactions
  of the ASME}, 125(4):709--715, 2003.

\bibitem{chuang2007}
J.-J. Chuang and D.C.H. Yang.
\newblock A laplace-based spiral contouring method for general pocket
  machining.
\newblock {\em The International Journal of Advanced Manufacturing Technology},
  34(7-8):714--723, 2007.

\bibitem{cormen2009}
T.H. Cormen, C.E. Leiserson, R.L. Rivest, and C.~Stein.
\newblock {\em Introduction to Algorithms}.
\newblock MIT Press, 2009.

\bibitem{dijkstra1959}
E.W. Dijkstra.
\newblock A note on two problems in connexion with graphs.
\newblock {\em Numerische Mathematik}, 1(1):269--271, 1959.

\bibitem{graham1983}
R.L. Graham and F.F. Yao.
\newblock Finding the convex hull of a simple polygon.
\newblock {\em Journal of Algorithms}, 4(4):324--331, 1983.

\bibitem{handler1973}
G.Y. Handler.
\newblock Minimax location of a facility in an undirected tree graph.
\newblock {\em Transportation Science}, 7(3):287--293, 1973.

\bibitem{held2001}
M.~Held.
\newblock {VRONI}: An engineering approach to the reliable and efficient
  computation of {V}oronoi diagrams of points and line segments.
\newblock {\em Computational Geometry}, 18(2):95--123, 2001.

\bibitem{held2009topology}
M.~Held and S.~Huber.
\newblock Topology-oriented incremental computation of {V}oronoi diagrams of
  circular arcs and straight-line segments.
\newblock {\em Computer-Aided Design}, 41(5):327--338, 2009.

\bibitem{held2009}
M.~Held and C.~Spielberger.
\newblock A smooth spiral tool path for high speed machining of 2{D} pockets.
\newblock {\em Computer-Aided Design}, 41(7):539--550, 2009.

\bibitem{held2014}
M.~Held and C.~Spielberger.
\newblock Improved spiral high-speed machining of multiply-connected pockets.
\newblock {\em Computer-Aided Design and Applications}, 11(3):346--357, 2014.

\bibitem{huertas2014}
J.L. Huertas-Tal{\'o}n, C.~Garc{\'\i}a-Hern{\'a}ndez, L.~Berges-Muro, and
  R.~Gella-Mar{\'\i}n.
\newblock Obtaining a spiral path for machining {STL} surfaces using
  non-deterministic techniques and spherical tool.
\newblock {\em Computer-Aided Design}, 50:41--50, 2014.

\end{thebibliography}
\bibliographystyle{plain}

\newpage
\appendix
\section{An algorithm defining the movement of the wave}\label{appA}

Here we describe how we define the speed and time of the wave at every point in the diagram
$\VD(\PP)$ in a polygon $\PP$ with no island.
There might be other models giving equally good or even better
results, but we think this model has an advantage of being quite simple to implement.
Recall that for a node $n$ in the diagram $\VD$, $\ttimeO[n]$ is the time when the wave
reaches $n$ and $\speedO[n]$ is the speed with which it passes $n$.
We define the movement of the wave from the root and out towards the leafs in
$\VD$.
We let $\ttimeO[\rootNode]=0$ and $\speedO[\rootNode]=\height[\rootNode]$.
Let $e$ be an edge going out of the node $n$ where we have
defined $\ttimeO[n]$ and $\speedO[n]$ already. Let $\pi$ be
the longest path starting with $e$. By definition,
$\pi$ has length $h=\ell[e]+\height[\nodeEnd[e]]$, where $\ell[e]$ is the length of $e$, i.e.,
the distance from node $n$ to node $\nodeEnd[e]$.
If $h<\height[n]$, the wave has to slow down on $e$, since the speed of the wave at node $n$
is determined by $\height[n]$.

We decrease the speed linearly
as a function of time such that the wave is decreasing on the first 1/4 of $\pi$
while it has constant speed on the last 3/4 of $\pi$.
The resulting spiral looks wrong if the wave abruptly changes acceleration when it is not needed.
Therefore, if the wave is already slowing down when reaching the node $n$, we might prefer that it keep
slowing down on $e$ with the same rate, even though we must use more than 1/4
of $\pi$. We do that if $\height[n]\leq 1.1\cdot h$, i.e., if $\pi$ is almost as long
as the longest path going out of $n$.
How to implement this idea is described in greater detail in the following.
It would be interesting to find a model where the acceleration is a continuous function
of the time along every path in $\VD$, but we have not found such a model that could be implemented
efficiently.

We define the values $\ttimeT[e]$ and $\speedT[e]$
for each edge $e$ which satisfy
$\ttimeT[e]\geq\ttimeO[n]$ and $\speedT[e]\leq\speedO[n]$,
where $n$ is the node $\nodeStart[e]$.
At a time $\ttimeO[n]<t<\ttimeT[e]$,
the speed of the wave is
$$(1-x)\cdot \speedO[n] + x\cdot \speedT[e],$$
where
$x = \frac{t-\ttimeO[n]}{\ttimeT[e]-\ttimeO[n]}$. When $t\geq\ttimeT[e]$, the speed
is $\speedT[e]$. Let $\getSpeed(e,t)$ be the speed defined by the values of edge $e$ and
$n=\nodeStart[e]$
at time $t$. Let $\getDist(e, t)$ be the distance travelled by the wave from time
$\ttimeO[n]$ to $t\geq \ttimeO[n]$.
We have
$$\getDist(e,t) = \int_{\ttimeO[n]}^t \getSpeed(e,u)du,$$
which can be computed easily since $\getSpeed(e,u)$ is piecewise linear.
We also need the function $\getTime(e,d)$ which is the time $t$ such that $\getDist(e,t)=d$.
Hence, $\getTime$ is the inverse of $\getDist$, i.e., we have
$\getTime(e,\getDist(e,t)) = t$ and $\getDist(e, \getTime(e,d)) = d$.
When $0<d<\getDist(\ttimeT[e])$, we can compute
$\getTime(e,d)$ by solving a quadratic equation. When
$d\geq \getDist(\ttimeT[e])$, we get a linear equation.
Finally, $\getPoint(e,t)$ returns the point
$(1-x)\cdot \point[n] + x\cdot \point[m]$,
where $x=\getDist(e,t)/\ell[e]$ and $m=\nodeEnd[e]$.
For $\ttimeO[n]\leq t\leq\ttimeO[m]$, $\getPoint(e,t)$ is the position of the wave on
edge $e$ at time $t$.

Assume that we have defined $\ttimeO$, $\speedO$, $\ttimeT$, and $\speedT$ on all
nodes and edges on the path from $\rootNode$ to some non-leaf node $n$.
Algorithm \ref{setTimesAndSpeeds} computes the values for an edge $e$ going out of
$n$ and for the node $\nodeEnd[e]$.

\begin{algorithm}[h]
\scriptsize
\LinesNumbered
\DontPrintSemicolon
\SetArgSty{}
\SetKwIF{If}{ElseIf}{Else}{if}{}{else if}{else}{end if}
\SetKwFor{Foreach}{for each}{}{end for}
\SetKwInput{Effect}{Effect}
$n\gets\nodeStart[e]$; $m\gets\nodeEnd[e]$; $t_n\gets \ttimeO[n]$; $v_n\gets \speedO[n]$\;
\uIf{$n=\rootNode$} {
  $t_e\gets \ttimeO[n]$; $v_e\gets \speedO[n]$\;
} \Else {
  $e_p\gets \parentEdge[n]$; $t_e\gets \ttimeT[e_p]$; $v_e\gets \speedT[e_p]$\;
}
$\ttimeT[e]\gets t_e$\; \nllabel{stasSetttimeT}
$\speedT[e]\gets v_e$\; \nllabel{stasSetspeedT}
$h\gets \height[m]+\ell[e]$\;
\If {$\getDist(e,1)>h$}
{ \nllabel{stasIf1}
  $\reuseAcc\gets \texttt{False}$\;
  \If {$t_n < t_e$ and $\height[n]\leq h\cdot 1.1$} { \nllabel{stasIf2}
    $a\gets \frac{v_e-v_n}{t_e-t_n}$\; \nllabel{stasCompA}
    $v_1\gets v_n + a\cdot (1-t_n)$\;
    $s\gets \frac{v_n+v_1}2\cdot(1-t_n)$\; \nllabel{stasCompS}
    \If {$s\leq h$} { \nllabel{stasIf3}
      Define $\ttimeT[e]$ and $\speedT[e]$ such that $\getDist(e,1)=h$
      and $\frac{\speedT[e]-v_n}{\ttimeT[e]-t_n}=a$.\; \nllabel{stasCompEq1}
      $\reuseAcc\gets\texttt{True}$\;
    }    
  }
  \If {$\reuseAcc=\texttt{False}$} {
    Define $\ttimeT[e]$ and $\speedT[e]$ such that $\getDist(e,1)=h$
    and $\getDist(e,\ttimeT[e])=0.25\cdot h$.\; \nllabel{stasCompEq2}
  }
}
$\ttimeO[m]\gets\getTime(e,\ell[e])$\; \nllabel{stasCompttimeO}
$\speedO[m]\gets\getSpeed(e,\ttimeO[m])$\; \nllabel{stasCompspeedO}
\caption{$\setTimesAndSpeeds(e)$}
\label{setTimesAndSpeeds}
\end{algorithm}

In lines \ref{stasSetttimeT}--\ref{stasSetspeedT},
we try to use the same values for $e$ as for the previous edge $e_p$. If,
however, the length $h$ of the longest path starting with edge $e$ is smaller
than the longest of all paths going out of $n$, we are in the case of line \ref{stasIf1},
where we need the wave to slow down.
Lines \ref{stasCompA}--\ref{stasCompS} compute the distance $s$ that the wave will travel
if it continues to decrease speed with the same rate until time $1$.
We can only keep using the same acceleration if $s$ is smaller than $h$.
If we cannot
keep using the same acceleration or the speed of the wave is not decreasing at the node $n$,
we define the values in line \ref{stasCompEq1}
as previously described.
Both of the lines \ref{stasCompEq1} and \ref{stasCompEq2}
give two equations in the two unknowns $\ttimeT[e]$ and $\speedT[e]$. Each pair of equations
lead to a quadratic equation in one of the unknowns,
and we need to choose the unique meaningful solution.
We assign the time and speed values to every node and edge in linear time
by traversing $\VD$ once. Algorithm \ref{setAllTimesAndSpeeds} sets times and speeds for all
nodes and edges by traversing $\VD$ once.

\begin{algorithm}[h]
\scriptsize
\LinesNumbered
\DontPrintSemicolon
\SetArgSty{}
\SetKwIF{If}{ElseIf}{Else}{if}{}{else if}{else}{end if}
\SetKwFor{Foreach}{for each}{}{end for}
\SetKwInput{Effect}{Effect}
  $(n_0,e_0)\gets (\rootNode,\childEdges[\rootNode][0])$\;
  $(n,e)\gets (n_0,e_0)$\;
  \Repeat{$(n,e)=(n_0,e_0)$} {
    \uIf {$n=\nodeStart[e]$} {
      $\setTimesAndSpeeds(e)$\;
      $n\gets \nodeEnd[e]$\;
    } \Else {
      $n\gets \nodeStart[e]$\;
    }
    $e\gets \getNextEdge(e,n)$\;
  }
\caption{$\setAllTimesAndSpeeds()$}
\label{setAllTimesAndSpeeds}
\end{algorithm}

\section{A method for defining time and speed on the central cycle}\label{appB}

In this section, we describe an alternative method to the naive one mentioned in section
\ref{defMovement} to define the time and speed of the wave on the cycle $\CC$
in the diagram $\VD(\PP\setminus\HH)$ of a polygon $\PP$ with an island $\HH$.

Recall that the \emph{preferred time} of a root node $n$ is
$$t_n=\frac{\holeHeight[n]}{\boundaryHeight[n] + \holeHeight[n]}.$$
We let the time for a node $n\in\CC$ be a weighted average of
its neighbors' preferred times.
We define an \emph{influence distance} $\inflDist[n]$
of each root node
$n\in\CC$, which is the distance along $\CC$ in which $n$ has influence on the times and
speeds of other root nodes. In many real-world instances, the majority of the trees
$\TT_n$ consist of just two edges, namely one going to $\PP$ and one to $\HH$, and these two
edges are almost equally long. We have experienced that these should have zero influence distance,
so that they only have an influence on their own times.
Therefore we give a positive influence distance if and only if one of the trees
$\HT_n$ and $\PT_n$ have more than one leaf or the ratio
$\frac{\holeHeight[n]}{\boundaryHeight[n]}$ is not in the interval $[1/1.02,1.02]$.

When the influence distance should be positive, we define it in the following way:
Consider three consecutive leafs $l_1$, $l_2$, and $l_3$ of $\VD$
on $\partial\PP$ or $\partial\HH$. We define the \emph{spanned boundary} of $l_2$
to be the path $[M_1,\point[l_2],M_2]$, where $M_1=\frac{\point[l_1]+\point[l_2]}{2}$
and $M_2=\frac{\point[l_2]+\point[l_3]}{2}$.
The spanned boundary of a tree $\PT_n$
is the union of all the spanned boundaries of the leafs of $\PT_n$, similarly for $\HT_n$.
We let $\inflDist[n]$ be the maximum of the distances between the start- and endpoints of the
spanned boundaries of $\PT_n$ and $\HT_n$.

To compute the times of the root nodes,
we define a weight of a root node $n\in\CC$ as
$\weight[n]=\holeHeight[n]+\boundaryHeight[n]+32\cdot\inflDist[n]$.
If a node $n$ with low weight or zero influence distance is very close (say, closer than $0.1\stepover$)
to a node $m$ with a high weight
(say, 5 times as much) and positive influence distance, we have experienced better results
if we set $\weight[n]=0$. In that way, the influential neighbor $m$ completely dominates node $n$.

Let $n$ be a fixed node on $\CC$
and consider another node $m$ on $\CC$ with positive influence distance.
Assume that the path
from $m$ to $n$ on $\CC$ has length $d\leq\inflDist[m]$.
We define the
weight of node $m$ on node $n$ as
$w_m=x_m^3\cdot\weight[m]$, where $x_m=1-\frac d{\inflDist[m]}$,
i.e., we let the weight decrease cubically as the distance increases.
The time at node $n$ is defined as
$$
\ttimeO[n]=\frac{\sum_m w_mt_m}{\sum_m w_m}.
$$
Here, the sums are over all nodes $m$ where $n$ is within the influence distance of $m$.

Recall that the \emph{preferred speed} of a root node $n$ is
$$v_n=\max\left\{\frac{\holeHeight[n]}{\ttimeO[n]},
\frac{\boundaryHeight[n]}{1-\ttimeO[n]}\right\}.$$
The wave should have a non-increasing speed from a root node $n$
towards the leafs of both $\HT_n$ and $\PT_n$.
Therefore, the speed at node $n$ should at least be the preferred speed
so that the wave can reach $\partial\HH$ at time $0$ and $\partial\PP$ at time $1$.
We define the speed as
$$
\speedO[n]=\max_m \left\{ v_m\cdot x_m^2\cdot (1-\lvert\ttimeO[n]-\ttimeO[m]\rvert)\right\},
$$
where the maximum is over all the nodes $m$ such that $n$ is within the influence distance of $m$.
The value $x_m$ is defined above.
The last factor in the expression is to reduce the influence from nodes that have gotten a
very different time than node $n$, since the speeds of the wave become less comparable
when the times are different.

\end{document}